\documentclass[manuscript, acmsmall, xcolor=svgnames,usenames,dvipsnames]{acmart}
\usepackage{algorithm}
\usepackage{algpseudocode}
\usepackage{tcolorbox}
\usepackage{xcolor}
\usepackage{float}
\usepackage{pifont}
\usepackage{pgfplots}
\usepgfplotslibrary{groupplots}
\def\norminf#1{\|#1\|_\infty}

\usepackage{listings}
\definecolor{gray}{rgb}{0.5,0.5,0.5}
\definecolor{mauve}{rgb}{0.58,0,0.82}
\definecolor{lightgrey}{rgb}{0.9,0.9,0.9}
\definecolor{darkgreen}{rgb}{0,0.6,0}

\newcommand{\NFMA}[0]{{N_{\mathrm{FMA}}}}

\newcommand{\neab}[0]{{n_{\mathrm{eab}}}}

\definecolor{matlabblue}{RGB}{0,0,255}        % keywords
\definecolor{matlabgreen}{RGB}{34,139,34}     % comments
\definecolor{matlabpurple}{RGB}{160,32,240}   % strings
\definecolor{matlabgray}{RGB}{128,128,128}    % line numbers / secondary
\definecolor{matlabbg}{RGB}{255,255,255}      % background (keep white for ACM)
\definecolor{mypink}{RGB}{255,105,180}

\newif\ifShowCorrections
\ShowCorrectionstrue
\ifShowCorrections
\usepackage[normalem]{ulem}
\definecolor{forgreen}{rgb}{0,0.6,0}
\definecolor{orange}{RGB}{255,140,0}
\definecolor{skyblue}{RGB}{100, 150, 235}

\newenvironment{mmc}{\begin{quote}\color{ForestGreen}\small\sf{$\clubsuit$~MM~}}{\end{quote}}

\newcommand{\swc}[1]{{\color{orange}[#1]}}
\else

\newcommand{\mmc}[1]{}

\newcommand{\swc}[1]{}
\fi

\AtBeginDocument{%
  }

\copyrightyear{none}
\copyrightyear{Version of: 11 June, 2026}

\makeatletter
\let\@copyrightfootnotetext\relax
\renewcommand\@copyrightpermission{}
\renewcommand\@formatdoi[1]{}
\makeatother

\begin{document}

\title{Accurate Models of NVIDIA Tensor Cores}

\author{Faizan A. Khattak}
\email{f.a.khattak@leeds.ac.uk}
\author{Mantas Mikaitis}
\email{mmikaitis@leeds.ac.uk}
\affiliation{%
  \institution{\\School of Computer Science, University of Leeds}
  \city{Leeds}
  \country{UK}
}

\renewcommand{\shortauthors}{Khattak and Mikaitis}

\begin{abstract}
  Matrix multiplication is a fundamental operation in both training of neural networks and inference. To accelerate matrix multiplication, Graphical Processing Units (GPUs) provide it implemented in hardware.
  Due to the increased throughput over the software-based matrix multiplication, the multipliers are increasingly used outside of AI, to accelerate various applications in scientific computing.
  However, matrix multipliers targeted at AI are at present not compliant with IEEE 754 floating-point arithmetic behaviour, with different vendors offering different numerical features.
  This leads to non-reproducible results across different generations of GPU architectures, at the matrix multiply-accumulate instruction level.
  To study numerical characteristics of matrix multipliers---such as rounding behaviour, accumulator width, normalization points, extra carry bits, and others---test vectors are typically constructed.
  Yet, these vectors may or may not distinguish between different hardware models, and due to limited hardware availability, their reliability across many different platforms remains largely untested.
We present software models for emulating the inner product behavior of low- and mixed-precision matrix multipliers in the V100, A100, H100 and B200 data center GPUs in most supported input formats of interest to mixed-precision algorithm developers: 8-, 16-, and 19-bit floating point.
These matrix multiplier models are first approximated by determining the numerical features via test vectors designed to trigger outputs sensitive to bit level differences in the implementation, followed by semi-exhaustive comparison (randomised input vectors of $10^7$ values) between the models and the actual GPU matrix multipliers---this process is repeated until the model is bit accurate.
These models enable verification of test vectors before applying them to real hardware and also support computational scientists and mixed-precision algorithm developers with easy-to-use accurate models available in MATLAB---we demonstrate their use in multi-word emulation algorithms for matrix multiplication.
\end{abstract}
% -------------

%   <concept_desc>Okay, Generate the Correct Terms for Your Paper</concept_desc>




\maketitle

\fbox{
  \parbox{5.3in}{The software associated with this paper, the \texttt{MATLAB Tensor Core v0.5}, which includes various NVIDIA GPU tensor core models as well as a generalised model that can be used to instantiate custom tensor core variants, is available on GitHub: \url{https://github.com/north-numerical-computing/MATLAB-tensor-core}.}
}

% ------------ Introduction-----
\section{Introduction}

Most recently released GPUs incorporate specialized matrix multiplier units, often referred to as \emph{tensor cores} or \emph{matrix cores},
which are designed to accelerate the GEneral Matrix Multiply (GEMM) operation for AI workloads.
Beyond neural networks, these units are also extensively used to accelerate fundamental linear algebra kernels used in high-performance computing (HPC) applications outside of AI~\cite{hbtd20, hima22, dgba25}. 
Although some applications may tolerate numerical inaccuracies, this is certainly not the case for all workloads. For example, tensor cores are increasingly utilized in applications such as Fourier transforms~\cite{fft_tc_1,fft_tc_2},~beamforming~\cite{tc_sp_2}, MR image reconstruction~\cite{tc_app_mri}, finite element simulations~\cite{finiteelemnt_tc}, high-precision dense matrix multiply emulation~\cite{mami25}, and various others~\cite{sc_tc_3,rounderr}.
At the same time, almost half of the machines on the November 2025 TOP500 list\footnote{\url{https://top500.org/lists/top500/list/2025/11/}} contain low-precision matrix multiplication operation in hardware (Fig.~\ref{fig:top500}).

Support for very low-precision, 8-, 6-, and even 4-bit, floating-point formats (albeit with higher precision accumulation in the inner products) is widely available in AMD~\cite{amd25} and NVIDIA~\cite{nvid25b} GPU architectures.
In this paper we will focus on NVIDIA matrix multipliers, informally known as \emph{tensor cores}.
Tensor cores generally do not adhere to any numerical standards, likely due to the absence of mixed-precision arithmetic standards at the time of their introduction in the NVIDIA V100 GPU in 2017.
Until the establishment of relatively recent Open Compute Project~\cite{modc23} and IEEE P3109~\cite{ieee25} standardisation activities, there was no standard for low-precision number formats and arithmetic.
Consequently, even today, AI hardware vendors implement mixed-precision matrix multipliers in various ways, and the computed matrix products across architectures and vendors may differ.
While some of the latest hardware implements the OCP data formats~\cite{modc23, rgsm23}, the OCP does not standardise arithmetic behaviour.

\begin{figure}
  \centering
      \begin{tikzpicture}
        \begin{axis}[
          ylabel={\# TOP500 machines},
          width=2.5in,
          thick,
          legend columns=3,
          legend style={at={(0.5,-0.5)},anchor=south,draw=none},
          x tick label style={rotate=45,anchor=east},
          /pgf/number format/.cd,
          use comma,
          1000 sep={},
          xtick={2016, 2020, 2025}
          ]

          \addplot[color=black, mark=diamond] table [x=year, y=16-bit] {data/gpus.dat};
          \addplot[color=OliveGreen!70, mark=triangle] table [x=year, y=matrix] {data/gpus.dat};
          \addplot[color=Fuchsia!70, mark=*] table [x=year, y=19-bit] {data/gpus.dat};
          \addplot[color=blue!70, mark=o] table [x=year, y=8-bit] {data/gpus.dat};

          \addplot[color=Red!70, mark=x] table [x=year, y=integer] {data/gpus.dat};

          \legend{16-bit FP, FP matrix mult., 19-bit FP, 8-bit FP, int matrix mult.};

        \end{axis}
      \end{tikzpicture}
      \caption{Number of machines on the November TOP500 lists that suppport low-precision floating-point formats, and low- and mixed-precision matrix multiplication operations. NVIDIA, AMD, and Intel GPUs are included in the counts.}
      \label{fig:top500}
\end{figure}

Mixed-precision tensor cores do not conform with the IEEE 754~\cite{ieee19} floating point standard, because that would require relatively expensive normalisation and rounding logic to be evaluated on every fused multiply-accumulate (FMA) operation within the matrix multiply-accumulate units, which may not be justified for AI workloads that tend to tolerate rounding errors.
The features of interest (that cause most impact to the matrix products) include normalization points in multi-term floating-point addition, the rounding modes at various points of computation, the block FMA size (number of additions in a dot-product-add operation), and the use of extra bits in the intermediate alignments of significands of products, amongst other design decisions.
The lack of standardization across vendors makes reproducibility and explainability of numerical results particularly challenging in scientific computing applications which may rely on cross-platform consistency.
It can also cause difficulties for performing accurate rounding error analysis of algorithms that utilise mixed-precision matrix multipliers.

To address these challenges, the identified numerical features can be abstracted into bit-accurate behavioral models that capture the characteristics of matrix multipliers. These models provide a software-level representation of hardware behaviour, enabling architecture research without requiring direct access to proprietary GPU implementations or specialised hardware knowledge. Furthermore, such models allow reasoning about the numerical accuracy of scientific computing applications that utilise matrix multipliers on data centre GPUs.

In addition, these models may help improve numerical reproducibility and stability in large-scale numerical simulations~\cite{rounderr}. Undocumented hardware-level numerical effects that contribute to variability in machine learning systems~\cite{random_ML_review}, once characterized, can help isolate true algorithmic behavior from hardware-induced artifacts, thereby improving experimental consistency. Moreover, since matrix multiplication is a core computational component of large language model inference, variations in GPU numerical behavior can also lead to differences in inference outputs across hardware platforms~\cite{inference_LLM_numerical}.
The proposed models make it possible to reproduce and modify such numerical behaviors in software, enabling controlled studies of their impact on the accuracy, and reproducibility.

\subsection{Previous work}

The numerical characteristics of mixed-precision matrix multipliers are seldom documented.
Nonetheless, recent studies have attempted to characterize some of these features using input vectors, compiled manually or through theorem provers, that trigger special cases in the matrix multipliers of different designs to produce unique results, allowing one to reason about the overall numerical behaviour without checking all the possible inputs.
Such studies have been done for AMD, and NVIDIA GPUs~\cite{hibr19, fhmp21,llfs24, vlpg25, khmi25}.

These works have developed conceptual models of matrix multipliers, but we argue, using randomised testing of tensor cores and their models, that refinement is needed to improve their accuracy.
The underlying limitation of these prior approaches is that they rely on an assumed feature space and subsequently determine whether such features are present in a given architecture. If a feature lies outside the assumed feature space, these methods are inherently unable to uncover it since no test is created to target it. In contrast, randomized testing compares model predictions against real hardware results over selected ensembles of inputs, enabling the discovery of hidden behaviours that may not belong to the originally assumed feature space, allowing to create a feedback loop into the testing methodology.

The goal of this project is to develop accurate MATLAB models of matrix multipliers, refined through a combination of targeted feature testing and randomized testing against GPU-generated results, to enable the mixed-precision research community to perform accurate numerical experiments when developing new algorithms that target these mixed-precision units.
We extend the past approaches and also embed them within an iterative technique that allows to refine the models after the initial approximation is obtained.

This type of work goes back to the \emph{Paranoia}\footnote{\url{https://www.arithmazium.org/paranoia/aaapara_toc.html}} software, built in the 1980s, for testing the compliance with the IEEE 754 standard~\cite{ieee19}.
Several projects followed \emph{Paranoia}.
Hillesland and Lastra~\cite{hila04} developed a GPU Paranoia which allowed them to analyse R300 and NV30 devices and found that, for example, basic arithmetic operations were not optimally accurate as prescribed by the \emph{correct rounding} of IEEE 754.
The FPGA Paranoia of Tan, Boland, and Constantinides~\cite{tbc12} similarly tested the arithmetic of various FPGA devices, finding for example that Altera devices rounded division differently from Xilinx devices, by not rounding to the nearest number in some cases.

\subsection{Contributions}

In this work, we reapply the generalized test vectors proposed in our earlier study~\cite{khmi25} to determine the numerical features of NVIDIA’s V100, A2, A30, A40, A100, H100, H200, L40S, Ada RTX 1000, RTX PRO 6000 and B200 tensor cores across all supported input and output precisions---40 models in total.
The GPU devices we target, along with their corresponding architectures and streaming multiprocessor (SM) compute capabilities, are summarized in Table~\ref{table:gpus}.
Unlike prior analyses \cite{hibr19, fhmp21, llfs24, vlpg25, khmi25}, we extend the investigation to include the two 8-bit floating-point formats available on the L40S, Ada RTX 1000, H100, H200, RTX PRO 6000, and B200 GPUs. Based on the identified features, we develop MATLAB-based software models of tensor cores for the aforementioned GPUs, for increasing the productivity and accessibility to tensor cores by mixed-precision algorithm developers and numerical analysts.
Features that we cannot find initially are identified by comparing the outputs of the emulated models against results from nine GPUs, and subsequently verified using targeted test vectors.
This iterative approach refines the MATLAB models to a higher level of accuracy than before.

\begin{table}[t]
  \begin{center}
  \caption{NVIDIA GPUs whose models of matrix multipliers are determined in this work, together with their corresponding microarchitecture and SM (streaming multiprocessor) compute capability.}\label{table:gpus}
  \begin{tabular}{llc}
    \toprule
    GPU Model & Architecture & SM Compute Capability \\
    \midrule
    V100 & Volta & 70\\
    A100& Ampere & 80 \\
    A30 & Ampere & 80\\
    A40 & Ampere & 86 \\
    A2 & Ampere & 86\\  
    L40S & Ada Lovelace & 89\\
    RTX 1000 & Ada Lovelace & 89 \\
    H100 & Hopper & 90 \\
    GH200 & Hopper & 90 \\
    B200 & Blackwell & 100 \\
    RTX PRO 6000 & Blackwell & 120 \\

    \bottomrule
  \end{tabular}
  \end{center}
\end{table}

In addition, our MATLAB toolbox provides a customizable tensor core model that allows users to diverge from one of the GPU tensor core models by customizing the precision and rounding behaviour.
The models are developed by combining fixed-point arithmetic with the custom floating-point format simulator CPFloat~\cite{fami23}.
We also demonstrate the use of the models for the emulation of high-precision matrix multiplication of arbitrary-sized dense matrices via tensor cores~\cite{mclp18, pili21, ooyo22, mami25}.

The main contributions of this paper are summarized as follows:
\begin{itemize}
\item Determination of various numerical features for NVIDIA V100, A100, A2, A30, A40, H100, H200, B200, L40S, RTX PRO 6000 and RTX 1000 Ada tensor core models, including 8-bit floating-point formats.
  We have presented these features more precisely than any other previous work, using architectural diagrams.
\item Development of MATLAB-based simulation models for GPU tensor cores, validated against GPU results through randomised matrix multiplication input space coverage.
\item A customised model of tensor cores in MATLAB, which can be used to easily experiment with custom variants by specifying various features such as rounding modes and internal precision used within tensor cores.
    \item Experimental results on the differences that tensor cores can cause in a low-level kernel, multi-word matrix multiplication, which serves as an example of how these models can be used by the community.
\end{itemize}

%%%%%%%%%%%%%%%%%%%%%%%%%%%%%%%%%%%%%%%%%%%%%%%%%%%%%%%%%%%%
\section{Notations and definitions}

Table~\ref{table:fp-formats} shows some of the characteristics of various floating-point formats available on the latest NVIDIA Blackwell architecture.
Hereafter we refer to floating-point formats with the following short-hand names: fp8 (either of fp8-E4M3 or fp8-E5M2), fp16 (binary16 of IEEE 754~\cite{ieee19}), bf16 (bfloat16), tf19 (TensorFloat32), fp32 (binary32 of IEEE 754), and fp64 (binary64 of IEEE 754).

\begin{table}[t]
  \begin{center}
  \caption{Floating-point formats that are available on the latest NVIDIA Blackwell GPUs~\cite{nvid25b}. Precision of the significand, which includes an implicit bit~\cite{ieee19}, minimum representable positive normalised value, and an approximate maximum representable positive value, are shown for each format.}\label{table:fp-formats}
  \begin{tabular}{lrlrlc}
    \toprule
    Format & precision & min norm. pos. & max pos. \\
    \midrule
    binary64~(double) & 53 & $2^{-1022}$ & $\sim 1.798 \times 10^{308}$  \\
    binary32~(single) & 24 & $2^{-126}$ & $\sim 3.403 \times 10^{38}$ \\
    tf19 (19-bit) & 11 & $2^{-126}$ & $\sim 3.401 \times 10^{38}$ \\
    bfloat16 & 8 & $2^{-126}$ & $\sim 3.389 \times 10^{38}$ \\
    binary16~(half) & 11 & $2^{-14}$ & $65504$ \\
    fp8-E4M3 & 4 & $2^{-6}$ & $448$ \\
    fp8-E5M2 & 3 & $2^{-14}$ & $57344$\\
    fp6-E2M3 & 4 & $2^{0}$ & $7.5$\\
    fp6-E3M2 & 3 & $2^{-2}$ & $28$ \\
    fp4-E2M1 & 2 & $2^{0}$ & $6$  \\
    \bottomrule
  \end{tabular}
  \end{center}
\end{table}

Take two matrices $A \in \mathbb{R}^{m\times k}$, and $B\in\mathbb{R}^{k\times n}$.
The matrix multiply-accumulate (MMA) operation produces
\begin{align*}
    D=AB+C \in \mathbb{R}^{m\times n}.
\end{align*}
Denote with $d_{ij}$ the element at $i$th row and $j$th column of $D$.
We can express it as the inner product between the $i$th row of $A$ and $j$th column of $B$ as
\begin{align*}
    d_{ij}=\sum_{\ell=1}^{k}a_{i\ell}b_{\ell j}+c_{ij}.
\end{align*}
To focus on the inner product as an underlying operation, rather than on particular elements of $D$, we omit the subscripts for simplicity, and we have
\begin{align}
  \label{eq:inner}
    d=\sum_{\ell=1}^{k}a_{\ell}b_{\ell}+c=\sum_{\ell=1}^{k}p_{\ell}+c.
\end{align}

The operations of the type \eqref{eq:inner} are often approximated in hardware by employing a multi-term floating-point adder~\cite{mika24}.
In such adders, the significands of addends are aligned relative to the largest exponent either in a single global alignment or combination of local and global alignment steps~\cite{kamk19, hcry20}, and then subsequently added via a compressor or a tree of adders with a single rounding and normalisation step.
For such many-term dot products, we define the number of product terms $p_\ell$ added in one go as the size of the \emph{block fused multiply-accumulate}, following the naming convention used in the rounding error analysis of tensor cores by Blanchard~et~al~\cite{bhlm20}, denoted with $\NFMA$.
For instance, the tensor core in the A100 data center GPU has $\NFMA=8$ for fp16 inputs which means $p_1+\dots+p_8+c$ is computed with a single normalisation and rounding.
When such units perform multi-term significand alignment, the bits of each significand that are shifted to the right are truncated or rounded with multiple sticky bits~\cite{tenc09}.
In the alignment of significands, we denote the number of extra alignment or guard bits beyond the output precision by $\neab$, similar to our previous work~\cite{khmi25}.
For instance, it has been shown by several studies that V100 GPU tensor core performs accumulation in 23 fractional bits (the precision of the fp32 format), hence with $\neab=0$~\cite{hibr19, fhmp21, llfs24}.

%We denote input and output precision as $\pin$ and $\pout$.

Note that below, we use some abbreviations which are specific to CUDA and PTX ISA.
Some of them are \texttt{HMMA}, \texttt{QMMA}, \texttt{DMMA}, \texttt{QGMMA}, \texttt{UTCHMMA}, \texttt{UTCQMMA} with meanings: matrix multiply and accumulate, fp8 matrix multiply accumulate, matrix multiply accumulate, fp8 matrix multiply and accumulate across a \emph{warp group} (a group of parallel threads), uniform matrix multiply and accumulate, uniform matrix multiply and accumulate, respectively~\cite{nvidia_instrset}.

%%%%%%%%%%%%%%%%%%%%%%%%%%%%%%%%%%%%%%%%%%%%%%%%%%%%%%%%%%%%
\section{Methods}

\subsection{Generalised Numerical Feature Testing (GNFT)}
This method for determining numerical features operates on the principle of using carefully designed expressions for forming test vectors that can trigger bit-level differences in the outputs~\cite{khmi25}.
For example, consider a test vector designed to determine whether subnormal inputs are supported in the fp16 format~\cite{ieee19}.
A test vector can be constructed by using constants: $|a_1| < 2^{-14}$ (smallest normalised value in fp16; see Table~\ref{table:fp-formats}), $b_1 = 1$, and $a_\ell = b_\ell = 0$ for $\ell = 2, \dots, k$.
This can be generalised for any format, by using an expression for the smallest normalised value.

When applying this test vector to a tensor core, if the resulting output $d$ is non-zero, it indicates that subnormal inputs are supported---an outcome that has to be formulated by understanding the IEEE 754 floating-point arithmetic.
Similarly, $a_1=2^{-14}$ and $b_1=2^{-1}$ would demonstrate if subnormals can be produced by arithmetic operations from normalised values.
Several such numerical feature–specific test vectors using constant values have been proposed in the literature~\cite{fhmp21,llfs24,hibr19}.
A generalized formulation, which addresses the limitations of earlier approaches by avoiding manually deriving constants or rerunning a theorem prover that has no upper bound on the run time~\cite{vlpg25}, for each format/device, has been explored by us~\cite{khmi25}---we rely on that approach here to determine the initial approximation of numerical features of tensor cores in eleven NVIDIA GPU variants and develop accurate models of them.

\subsection{Input Space Search Method (ISSM)}

Once the GNFT method has identified the numerical features, a conceptual or software-based model of the hardware matrix multiplier can be constructed to simulate its behavior.
To achieve higher confidence in this model, in comparison to the GPU implementation, the ISSM method is invoked to look for input vectors to \eqref{eq:inner} for which the GPU results deviate from those produced by the software model approximated by GNFT.
Such discrepancies can then be analyzed to refine the model until the tests pass.
When mismatches occur, a generalised test vector can be constructed to specifically detect and characterize such numerical features when analysing a new hardware. This expands the feature space and therefore, simplifies the process of determining new architectures.

It is worth to note that the input space of \eqref{eq:inner} is generally large and only a small proportion of it can be checked.
For example, for $k=16$ and 8-bit floating-point as an input format, the input space pair $\{a,b\}$ has approximately $256^{32}\approx 10^{77}$ possible inputs.
For this work we have chosen to sample $a,~b$, and $c$ from a standard normal and uniform distributions along with carefully designed cases which is detailed in Section 4.2.

For each of the GPUs we have tested, some of them have up to five input formats: fp8 (both E4M3 and E5M2), fp16 (both fp16 and fp32 output mode), bf16, and tf19, requiring a separate verification for each. 
In total, we have verified 11 models of tensor cores, each with several input/output format combinations, giving in total 43 model verification runs with randomized input vectors (see Table~\ref{tab:nf}).

\subsection{Matrix multiplier model approximation and refinement}
\label{sec:main-alg}

Our proposed algorithm for determining an accurate model of a tensor core is shown in Algorithm~\ref{alg:refine-model}.

\begin{algorithm}
\caption{Pseudocode for approximation and refinement of models of matrix multipliers.}
\label{alg:refine-model}

\begin{algorithmic}[1]

\State \textbf{Machine step:}
Apply GNFT using generalised test vectors~\cite{khmi25}.
\State \textbf{Engineer step:}
Create an initial approximation of the model.

\While{true}

    \State \textbf{Machine step:}
    Generate an ISSM ensemble as described in Section~4.2 using~Algorithm~\ref{alg:ensemble_genertion}.

    \State \textbf{Machine step:}
    Evaluate both the GPU and the approximate model over the entire ensemble.

    \If{mismatches are detected between GPU and model outputs}

        \State \textbf{Engineer step:}~Inspect failure cases.
        \State \textbf{Engineer step:}~Refine the model.
        \State \textbf{Engineer step:}~Modify ISSM to include similar cases to the failure cases.
        \State \textbf{Engineer step}~Add test expressions to GNFT for detecting any new features.

    \Else

        \State \textbf{Engineer interpretation:}
        The model is considered sufficiently accurate.

        \State \textbf{break}

    \EndIf

\EndWhile

\end{algorithmic}
\end{algorithm}

The step that modifies the model (line 8) causes the biggest challenge in automating the whole process, because it requires an expert to look into the mismatching outputs between the GPU and the model, and develop hypotheses about the features of the model that need to be changed to match the GPU output.
Through Algorithm~\ref{alg:refine-model} a lot of the manual work is removed by using the generalised testing vector to get the first approximation of the model and refinement, which means that Algorithm~\ref{alg:refine-model} can significantly accelerate the determination of features of future tensor cores.
However, the full automation of Algorithm~\ref{alg:refine-model} is an open problem.

%%%%%%%%%%%%%%%%%%%%%%%%%%%%%%%%%%%%%%%%%%%%%%%%%%
\section{Results}

%%%%%%%%%%%%%%%%%%%%%%%%%%%%%%%%%%%%%%%%%%%%%%%%%%%%%%%%%%%
\subsection{Accurate GPU Matrix Multiplier Models}
Numerical features of matrix multiplier units on AMD and NVIDIA have been determined up to a certain degree of accuracy~\cite{hibr19, fhmp21, llfs24, vlpg25}. Focusing on NVIDIA GPUs, we argue that previous work essentially corresponds only to the first step of Algorithm~\ref{alg:refine-model}, albeit not using generalised tests but constant-based format-specific expressions, and that the resulting models remain inaccurate without further refinement.
Previous approaches require manual porting of tests in line 1 of Algorithm~\ref{alg:refine-model} for each new GPU/format combination, but with our GNFT~\cite{khmi25} this limitation is removed.

The numerical features of all eleven GPU tensor cores are summarized in~Table~\ref{tab:nf}, including the product alignment bits, accumulator output precision, $\NFMA$, final rounding mode to output precision, and other relevant details.

\subsubsection{V100 Data Center GPU}

\begin{figure}[!t]
  \centering
  \includegraphics{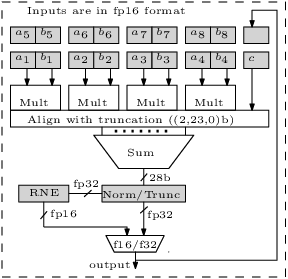} % replace with your file
  \caption{A model of the inner product within the V100 GPU tensor core. Here, RNE denotes round-to-nearest ties-to-even~\cite{ieee19} rounding mode.}
  \label{fig_V100_model}
\end{figure}

Mixed-precision matrix multiplication units, or tensor cores, were introduced in the NVIDIA Volta architecture, namely, the V100 device.
The V100 tensor cores support only fp16 as the input format with fp16 or fp32 as the output format~\cite{nvid17}.
Xi~et~al.~\cite{llfs24} claim that the FMA size cannot be determined when there are no extra alignment bits compared with the precision of the output format.
Following~\cite{khmi25} we denote this situation with $\neab = 0$.
However, using the FMA size, $\NFMA$ (detection algorithm proposed in~\cite{khmi25}), we were able to identify $\NFMA=4$, which we could determine irrespective of the value of $\neab$.
In addition, when the significands are aligned, bits that fall outside the internal accumulator's precision are truncated rather than accumulated into multiple sticky bits as is done in the algorithm by Tenca~\cite{tenc09}.
Further tests on the V100 GPU using this approach led us to the estimated tensor core model shown in Fig.~\ref{fig_V100_model}, which differs from the model presented in~\cite[Fig.~2]{inte18} in that the significand alignment is performed in 25 bits rather than 24, and $c_1$ is added together with the products rather than separately.

In our experiments, the existent conceptual model, when simulated in MATLAB, produced different results from the GPU for certain input combinations in the inner loop of Algorithm~\ref{alg:refine-model}.
Upon analysis, we determined that in some cases, partial products are not normalized before accumulation, which can result in variations even for apparently identical products.
For instance, consider the case with $c = 0$, $p_1 = 2.25$, and $p_2 = p_3 = 2^{-23}$.
Now consider two alternative ways to obtain the product $p_1 = 2.25$: either by taking $a_1 = b_1 = 1.5$, or $a_1 = 1$, $b_1 = 2.25$.
For the first case, the tensor core computed $d = 2.25 + 2^{-22}$, whereas for the second case, $d = 2.25$.
In the first case, $p_1$, which has the largest magnitude, is added in a denormalised form ($10.01_2 \times 2^0$), and hence $p_2$ and $p_3$ are not truncated.
In contrast, in the second case, since $p_1 = 01.001_2 \times 2^1$, the remaining two products $p_2$ and $p_3$ fall outside the representable range (beyond the 23rd fractional bit), resulting in $d = 2.25$.
On the other hand, if $c=2.25$ with $p_1=0$, $p_2=p_3=2^{-23}$, we have $d=c$.
This feature is depicted in Fig.~\ref{fig_V100_model} as $(2, 23, 0)$ bits at alignment stage where $2$ represents integer bits, $23$ fractional bits, and $0$ the $\neab$ bits.
It is worth noting that this test can be generalized, similar to our previous work~\cite{khmi25} as it does not rely on the specific value such as \(2.25\); rather, any product greater than \(2\) and less than \(4\) can be used where the corresponding $a$ and $b$  have values $2>a,b\ge \sqrt{2}$.
%This special case does not occur if $|c|\ge |p_\ell|~\forall\ell$.
One possible explanation for this behaviour is that it improves numerical accuracy because the products that need normalising do not lose one bit.
This feature also means that the products may be allowed to exceed $2^{128}$ (not representable in fp32; see Table~\ref{table:fp-formats}) and be used in the subsequent summation.
This is not possible to be tested with the fp16 input format where the maximum product exponent remains capped at $30$ for denormalised and $31$ for normalised products.

\subsubsection{A100 Data Center GPU}
\label{sub_sec_A100}
Now we turn to several tensor core variants in the A100.

\begin{figure}[!t]
  \centering
  \includegraphics{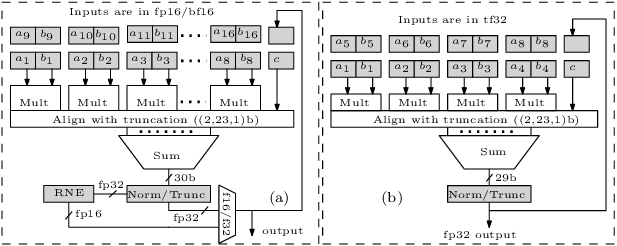}
  \caption{A model of the inner product within the A100 GPU tensor core for the three input formats. A100 also has an fp64 tensor core, but that tensor core is compliant with the IEEE 754 FMA operation and is not shown here.}
  \label{fig_A100}
\end{figure}

In the fp32 output mode with fp16/bf16 as the input format, $\NFMA=8$, and $\neab=1$ is used for the alignment and accumulation of significands of products $p_{\ell}$.
The rounding mode at both the alignment and post-normalization stages is truncation, whereas in fp16 output mode the rounding mode after normalization is Round-to-Nearest Ties-to-Even (RNE).
The alignment of the significands is $26$ bits wide, with $2$ integer bits and $24$ fractional bits, while the adder output must be 
\[
26 + \left\lceil \log_{2}(9) \right\rceil = 30 \ \text{bits}
\] 
due to extra bits required for accumulation carries.
We note that only three extra carry bits can be detected through testing~\cite{khmi25}, and although a fourth carry bit is assumed to exist by logic, to the best of our knowledge it cannot be verified through numerical tests unless we assume that there is no intermediate normalisation~\cite{vlpg25} when computing \eqref{eq:inner}.\footnote{A test vector is needed, that can distinguish normalisation followed by the addition of the final addend, versus addition of the final addend to an accumulator that is in a denormalised form with the fourth carry bit set, followed by the final normalisation and truncation instead of rounding.}
The model that we have determined is shown in Fig.~\ref{fig_A100} (a).

For the tf19 input mode, the $\NFMA=4$, with truncation as the default rounding mode.
The model is shown in Fig.~\ref{fig_A100} (b).
The bit-width of the denormalised sum must be $29$-bit wide.

The products remain denormalised for fp16/bf16/tf19 input modes, as previously discussed in the context of V100 and fp16. This behaviour can be verified by setting
$c = 0$, $p_1 = 2.25$ where $a_1 = b_1 = 1.5$, and $p_2 = 2^{-23}$, $p_3 = p_4 = 2^{-24}$.
When $d = 2.25 + 2^{-22}$, the products are added in denormalised form.
The difference in this test compared to the V100 lies in the presence of an extra alignment bit
($\neab = 1$) in the A100, which slightly alters the accumulation behaviour, making this test dependent on knowing the $\neab$ value. 

There is a nontrivial numerical behaviour in bf16 and tf19 input formats, which we discovered by exploring the mismatch between our model and the GPUs, reported on GitHub by a user.
  This occurs within the alignment of product significands, when all product exponents fall within the subnormal range of the fp32 format.
Although product exponents are allowed to go as low as $-266$ for bf16 and $-272$ for tf19 inputs, the maximum exponent used for aligning significands during accumulation is capped from below; we denote this cap by $e^{\mathrm{align}}_{\mathrm{min}}$. We observed the following cases:

\begin{enumerate}
    \item when $c = 0$, and the product exponents are less than or equal to $-132$, the maximum exponent used for alignment is fixed at $e^{\mathrm{align}}_{\mathrm{min}}=-132$, even if the largest exponent among the products is smaller.  
    For example:
    \begin{itemize}
        \item setting $c = 0$, $p_1 = \sum_{i=-150}^{-155} 2^i$, $p_2 = p_3 = 2^{-156}$ results in $d = 2^{-149}$.
        \item for $c = 0$, $p_1 = \sum_{i=-150}^{-155} 2^i$, $p_2 = 2^{-156}$, $p_3 = p_4 = 2^{-157}$, we get $d = 0$.
    \end{itemize}
    This shows that $-132 + 156 = 24$ fractional bits are available on the A100 tensor core accumulator.
    In the second case, $25$ fractional bits are required to produce $2^{-149}$ as the output---truncation occurs and the result becomes zero.

    \item when $c$ is a nonzero subnormal value (i.e., $0 < |c| < 2^{-126}$) and all product exponents are in the fp32 subnormal range, the maximum exponent for alignment is set to $-126$ instead of $-132$. 
    This is verified by setting $c = 2^{-127}$, $p_1=2^{-150},p_2 = p_3 = 2^{-151}$ which produced $d=2^{-127}$ instead of $d=2^{-127}+2^{-149}$.

    \item when $c = 0$ and some or all product exponents are between $-132$ and $-126$, the maximum product exponent itself is used for alignment rather than the fixed value $-132$.
\end{enumerate}

In fp64 input/output mode, we have $\NFMA=1$.
This configuration of a tensor core is compliant with IEEE 754 with RNE as the default rounding mode, but supports RD, RU and RZ.
The observed accumulation order appears to follow
\[
((c + p_1) + p_2) + \dots,
\]
as indicated by permuting the values $1$, $2^{-53}$, and $2^{-53}$ among $c$, $p_1$, and $p_2$.
Only for the case $c = p_1 = 2^{-53}$ and $p_2 = 1$ does the GPU's tensor core produce $d = 1 + 2^{-52}$; for all other permutations, it returns $d = 1$ under the default RNE rounding mode.
This behavior further suggests that the FP64 tensor core performs sequential FMA operations, where each intermediate result is fed back as the new input $c$.

%% A2 A30
\subsubsection{A2 \& A30 Data Center GPUs}
Executing the GNFT on these GPUs and using the ISSM testing, we were able to determine that the tensor cores in these GPUs behave identically to the tensor cores of the A100 GPU, except that they do not have a fp64 variant.

%% Ada Lovelace
\subsubsection{L40S Data Center GPU Model}
\label{sec:l40s}

Running the GNFT for fp16, bf16, and tf19 input formats, we determined that the L40S tensor cores exhibit behaviour identical to that of the A100, with $\NFMA$ of $8$, $8$, and $4$, respectively, and with a single extra alignment bit i.e., $\neab=1$.

\begin{figure}[!t]
  \centering
  \includegraphics{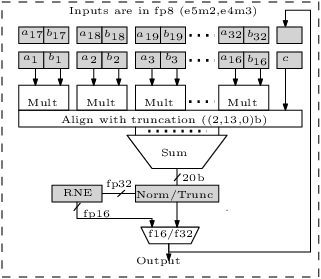}
  \caption{A model of the inner product within the tensor cores of the L40S and Ada Lovelace RTX 1000 GPU for the fp8 input format. For fp16, bf16 and tf19, the model is identical to A100 and is not shown.}
  \label{fig_L40S}
\end{figure}

The warp matrix multiply accumulate (WMMA) API does not provide support for fp8 format tensor cores available on this GPU.
Therefore, the MMA instruction, specifically
\path{mma.sync.aligned.m16n8k16/32.f32/f16.f8.f8.f32/f16}\footnote{The input matrices $A$ and $B$ may have a shared dimension of either $16$ or $32$. The output type can be either fp32 (denoted as \texttt{f32}) or fp16 (denoted as \texttt{f16}) in this PTX format.},
is utilized which then maps to \texttt{QMMA} instruction, for computing fp8 MMA across a warp~\cite[Sec.~6.4]{nvidia_instrset}. 
However, although the L40S GPU belongs to the Ada architecture family, the QMMA instruction is not explicitly documented in the Ada instruction set~\cite[Sec.~6.2]{nvidia_instrset}.
For both fp8 formats, fp8-E5M2 and fp8-E4M3, the accumulation takes place with $13$ fractional bits.
We determined this by setting $p_1 = 1$, $p_2 = p_3 = 2^{-23-\neab}$ and decrementing $\neab$ from $10$ until $d$ becomes equal to $1+p_1+p_2$, using the methodology of Hickmann~and~Bradford~\cite{hibr19}.
For $\neab= -10$, we obtain $d = 1 + 2^{-12}$, which indicates that there are $23-10 = 13$ fractional bits.
Moreover, we test that the accumulation term
$c$ is summed together with the products.
We verified it by setting $c = 1$ and $p_1 = p_{2} = 2^{-14}$, which produced $d = 1$ in the fp32 output mode.
As L40S fp16, bf16 and tf19 input format tensor cores are identical to A100 tensor core, only fp8 input format tensor core structure is shown in~Figure~\ref{fig_L40S}.
The A100 feature of limiting the maximum exponent at $-132$ when small subnormals are accumulated, is also available in this GPU.

\subsubsection{Ada Lovelace RTX 1000 Consumer-Grade GPU Model}
\label{sec:rtx1000}

Tensor cores on this GPU exhibited identical numerical characteristics to those of the L40S model, and produced identical outputs for $10^{5}~\text{and}~10^{7}$ random input vectors. 
The $e^{\mathrm{align}}_{\mathrm{min}}$ parameter is $-132$.

%-------------

\subsubsection{H100/H200 Data Center GPU}
\label{sec:h100}
For fp16/bf16 input formats, the H100 tensor core is reported by Li et al.~\cite{llfs24} to utilize two or more extra alignment bits, with fp32 as an output format, when the significands of the products are aligned for multi-term addition and summed.
Using our generalised test vectors~\cite{khmi25}, we confirm that the H100 actually employs exactly $\neab=2$ extra alignment bits; no third alignment bit is present, and the bits of aligned significands that are out of range are truncated for fp16/bf16/tf19 input formats.
As a result, the alignment stage is $26$-bits wide.
However, since the products are added in denormalised form, same as in the V100 and A100 devices, the width of the significand alignment is $(2,23, 2)=27$-bit wide, i.e., $2$ integer, $23$ fractional bits, and extra $2$ bits of the fraction.
This feature can be verified by setting $c=0$, $p_1=2.25$,
where $a_1=b_1=1.5$, $p_2=2^{-23}+2^{-24}+2^{-25}$, and $p_3=2^{-25}$.
If $d=2.25+2^{-22}$, the largest product is kept denormalised, otherwise the output should be $d=2.25$.

Furthermore, for the fp16 and bf16 input modes $\NFMA=16$, which equals the maximum supported shared dimension of the matrix sizes available in the WMMA API.
Since $\NFMA=16$, the denormalized adder result has a width of $27 + \left\lceil \log_{2}(17) \right\rceil = 32$ bits.
Upon normalization, results are truncated rather than rounded.
The fp16 format for both the input and output is supported, and the default rounding mode from the internal precision to the fp16 output is RNE.

\begin{figure}[p]
  \centering
  \includegraphics[scale=1.0]{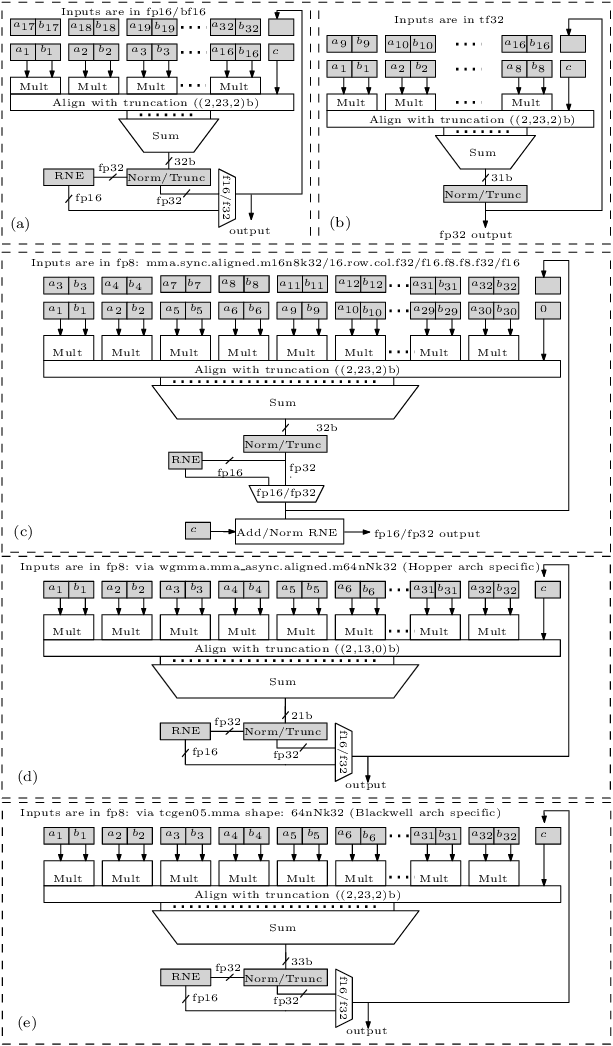}
  \caption{A model of the inner product within the tensor cores of the H100/H200/B200 GPUs for (a) fp16/bf16, (b) tf19, (c) fp8 input format provided via \texttt{mma.sync} (which internally uses fp16 tensor core with interleaved input pattern), (d) fp8 tensor core accessed via \texttt{wgmma.mma\_async} (specific to Hopper architecture), and (e) fp8 input format 5th generation tensor core in Blackwell architecture. 
  The fp64 tensor core is compliant with IEEE 754 FMA operation and is not shown.}
  \label{fig_H100}
\end{figure}

Using the WMMA API with the only available shape for the input matrix, i.e., \texttt{m16n16k8}, for tf19 input format, the instructions are mapped to \texttt{HMMA} but with fragment size of \texttt{m16n8k4} which results in $\NFMA=4$ via GNFT.
However, if \texttt{mma.sync.aligned} instruction is used with the matrix shape of \texttt{m16n8k8}, it is internally mapped to \texttt{HMMA.1688} which then yields $\NFMA=8$ through GNFT.
We report the maximum, $\NFMA=8$, in the model of H100 and H200 (see Figure~\ref{fig_H100}).
The  $e^{\mathrm{align}}_{\mathrm{min}}=\ -133$, unlike $-132$ in the case of Ampere and Ada architectures, when $c = 0$ is also present.
  However, the test case derived for the A100 is not directly applicable to H100/H200, as the tensor cores in the Hopper architecture have $\neab=2$.
  We now have the following.
\begin{enumerate}
\item when $c=0$, the maximum exponent used for alignment stage is limited at $-133$, as shown by
 \begin{itemize}
        \item setting $c = 0$, $p_1 = \sum_{i=-150}^{-156} 2^i$, $p_2 = 2^{-157},~p_3=p_4=2^{-158}$ results in $d = 2^{-149}$.
        \item for $c = 0$, $p_1 = \sum_{i=-150}^{-156} 2^i$, $p_2 = 2^{-157}+2^{-158}$,
          $p_3 = p_4 = 2^{-159}$, we get $d = 0$.
    \end{itemize}
    Since $\neab=2$, there are $25$ fractional bits i.e., $-133+158=25$, which prevents truncation of the $=p_3=p_4=2^{-158}$ term whereas the products less than $2^{-158}$ get truncated, producing $d=0$.
    \item when $c$ is a nonzero subnormal value, and all product exponents are in the fp32 subnormal range, the maximum exponent for alignment is set to $-126$. 
    This is verified by setting $c = 2^{-127}$, $p_1=2^{-151}$, $p_2 = p_3 = 2^{-152}$ which produces $d=2^{-127}$ instead of $d=2^{-127}+2^{-149}$, while $p_2=p_3>2^{-152}$ does not result in truncation.
\end{enumerate}

With WMMA API limited to fp16, bf16, tf19 and binary64 formats, \texttt{mma.sync.aligned} instruction can be used to multiply fp8 input format matrices, as in Ada RTX 1000 and L40S.
However, unlike Ada RTX 1000 or L40S where this is mapped to \texttt{QMMA}, in the case of H100/H200, we observed that fp8 inputs are first converted to fp16, after which the \texttt{HMMA} instruction is invoked at the SASS level\footnote{SASS is a low-level assembly language that compiles to binary microcode, which executes natively on NVIDIA GPU hardware, as per \url{https://archive.docs.nvidia.com/gameworks/content/developertools/desktop/ptx_sass_assembly_debugging.htm}} with a matrix shape of \texttt{m16n8k16}.
This indicates that a direct \texttt{QMMA} instruction is not available in the Hopper architecture, and fp8 inputs issued via \texttt{mma.sync.aligned} are effectively processed using fp16 tensor cores.
The internal usage of fp16 tensor core for fp8 input matrices results in interesting features for fp8.
We observed that the inner product is computed in interleaved fashion where 32-element input vectors are distributed across two tensor core invocations by alternating pairs of elements.
Once the products are added together,
$c$ is then added to the normalised sum of products, with RNE as the rounding mode (see Figure.~\ref{fig_H100}c).
This \emph{late addition} of $c$ is only present in this case, while in all other cases, $c$ is summed with the products.

The interleaved input pattern is detected in both cases for \texttt{m16n8k16} and \texttt{m16n8k32} matrix shapes.
It can be verified by fixing $p_1 = 1$ and $p_2 = 2^{-24}$, and then assigning the value $2^{-24}$ sequentially to $p_3$ through $p_{32}$ while keeping all other entries at zero. If $d = 1 + 2^{-23}$, the tested product falls into the same accumulation group as $p_1$ and $p_2$. Conversely, if $d = 1$, the tested product belongs to the second group. If the two groups were not interleaved, we would observe $d = 1 + 2^{-23}$ only when assigning $2^{-24}$ to positions $p_3$--$p_{16}$, and $d = 1$ when assigning it to positions $p_{17}$--$p_{32}$.

The Hopper architecture supports a specific \emph{warpgroup} level
\texttt{mma} instruction \texttt{wgmma.\allowbreak mma\_async.\allowbreak sync}, which is internally mapped to \texttt{QGMMA}, responsible for fp8 MMA across a warpgroup (a set of four contiguous warps)~\cite[Sec.~6.3]{nvidia_instrset}---the low-level instruction which supports native fp8 format tensor core access on Hopper architecture.
With this, the GNFT shows $\neab=-10$, i.e., $13$ fractional bits and $\NFMA=32$.
The behaviour of the fp8 format tensor core for H100/H200 is identical to that of the Ada Lovelace architecture, but with twice the $\NFMA$.
The model diagram for H100 and H200 is shown in Figure~\ref{fig_H100}d.

%----------------------------------------------------------
\subsubsection{B200 Data Center GPU Model}
\label{sub_sec_B200}
B200, based on the Blackwell architecture, is the latest iteration of the GPU architecture that is commercially available at the time of writing.
In our testing, the B200 tensor cores demonstrated identical numerical behaviour to the H100/H200 models (Fig.~\ref{fig_H100}) for 16- and 19-bit input formats.
For fp32 outputs, with fp16, bf16 or tf19 inputs, the accumulator provides $2$ extra bits, and the FMA tile sizes are $16$, $16$ and $8$, respectively. 
The product exponent limitation during significand alignment, when the product lies in the subnormal region, is identical to that observed on H100 and H200 GPUs.
When we tested the fp8 tensor core through the \texttt{mma.sync.aligned} PTX instruction, we observed an identical behaviour as that in the case of H100/H200, where fp8 input vectors are converted to fp16 and \texttt{HMMA} instruction is called at \texttt{SASS} level which means that fp16 tensor cores are utilized instead of fp8 cores.
Identical interleaving pattern is detected (see Figure~\ref{fig_H100}(c)).
Even though \texttt{QMMA} is specifically mentioned in the instruction set of Blackwell GPU~\cite[Sec.~6.4]{nvidia_instrset}, our experiments show that \texttt{mma.sync.aligned} instructions are mapped to \texttt{HMMA} not \texttt{QMMA} and \texttt{wgmma.mma\_async} is not supported.
This confirms that \texttt{QMMA} requires PTX instructions other than \texttt{mma.sync.aligned}.
Our experience is consistent with the findings of~ Jarmusch~et~al.~\cite{25dissectingnvidiablackwellarchitecture}.

To access the 5th generation tensor cores available on the B200, PTX instruction of \texttt{tcgen05.mma.\allowbreak cta\_group\_1/2::\allowbreak kind} is invoked where \texttt{kind} defines the input type and takes on values f16 (supporting both fp16 and bf16), tf32 (tf19), and f8f6f4 (all combinations of fp8, fp6, and fp4, of which the latter two we did not address in our work).
For \texttt{kind} set to f16/tf32, it maps to \texttt{UTCHMMA}, having the same numerical behaviour as that of the \texttt{HMMA}.
While for \texttt{kind} set to f8f6f4 with input in fp8 format we detect $\NFMA=32$ with $25$ fractional bits (i.e., $\neab=2$), the fp8 tensor cores in the previously discussed H100/H200 and Ada architectures provide only $13$ fractional bits (i.e., $\neab=-10$). 
The fp8 tensor core PTX instructions map to \texttt{UTCQMMA} in SASS.

\subsubsection{RTX Pro 6000 Workstation GPU (Blackwell)}

This GPU is based on the Blackwell architecture and is designed for desktop/workstation platforms. Similar to the B200, it features fifth-generation tensor cores.
The tensor core numerical features are identical to those of the B200; however, fp8 input format tensor core is accessed through the \texttt{mma.sync.aligned} PTX instruction, as in the Ada architecture, rather than the \texttt{tcgen05.mma} instruction used on the B200. Furthermore, it then internally maps these operations to \texttt{QMMA} instructions instead of \texttt{UTCQMMA} at SASS level.

Table~\ref{tab:cf} summarizes the CUDA/PTX instructions used in this work for feature extraction and randomized testing, together with their corresponding mappings to SASS instructions.

%%%%%%%%%%%%%%%%%%%%%%%%%%%%%%%%%%%%%%%%%%%%
%%%%%%%%%%%%%%
%\subsection{Other Potential Models}
%\subsubsection{IEEE 754 Standard Compliant Model}
%In contrast to the A100, V100, and H100 GPUs, which are not fully IEEE 754 standard compliant for fp16, BF16, and tf19 input modes, we propose a tensor core model that is compliant.In this mode, the FMA size is set to 1, and normalization is performed after every binary addition. Three extra alignment bits---two guard bits and one sticky bit---are used to implement all four rounding modes recommended by the IEEE 754 standard.The order of addition operations is kept similar to that of the A100/H100 tensor core model in fp64 format where $c$ is added early or along the products instead of late i.e., $((c+p_1)+p_2)+\dots$.
%\subsubsection{Custom Model}
%In the accompanying MATLAB code, we allow the user to simulate user-defined features, simulating custom tensor core models---FMA size, desired number of extra alignment bits and rounding modes can be modified. In addition, one can allow infinite precision during the accumulation of multi-term floating point addition by setting the extra alignment bits to Inf.

%----------------------------------

\begin{table*}
  \begin{center}
  \caption{Summary of the numerical features of several generations of  NVIDIA tensor cores (mixed-precision matrix multipliers) in nine different GPUs spanning years of release from 2017 to 2024.}\label{tab:nf}
  % Second version of table, with booktabs.
{     
\begin{tabular}{lllccccc}\toprule

  In & Out & GPUs & Prd. Align. & Acc. & $\NFMA$ & $e^{\mathrm{lim}}_{\mathrm{min}}$ & Final Round. \\\midrule
fp8 & fp32 & B200, RTX PRO & (2,25) & 33 & 32 & - & Trunc.  %early 
\\
& &
H100, H200 & (2,13) & 21 & 32 & - & Trunc  %early 
\\
         & & L40S, Ada RTX 1000 & (2,13) & 20 & 16 & - & Trunc %early
         \\

fp8 & fp16 & B200, RTX PRO & (2,25) & 33 & 32 & - & RNE %early
\\ 
& & H100, H200  & (2,13) & 21 & 32 & - & RNE %early
\\
         & & L40S, Ada RTX 1000 & (2,13) & 20 & 16 & - & RNE  %early 
         \\

%fp8* & fp16/fp32 & H100, H200, B200 & (2,25) & 32 & 16 & RNE & late \\

fp16/bf16 & fp32 & H100/H200/B200/RTX PRO & (2,25) & 32 & 16 & $-133$ & Trunc  %early
\\
                  & & L40S, Ada RTX 1000 & (2,24) & 30 & 8 & $-132$ & Trunc  %early
                  \\
                  & & A100, A2, A30 & (2,24) & 30 & 8 & $-132$ & Trunc  %early
                  \\
                  & & V100 & (2,23) & 28 & 4 & - & Trunc  %early
                  \\

tf19 & fp32 & H100/H200/B200/RTX PRO & (2,25) & 31 & 8 & $-133$ & Trunc %early 
\\
          & & L40S, Ada RTX 1000 & (2,24) & 29 & 4 &$-132$ & Trunc % early 
          \\
          & & A100, A2, A30 & (2,24) & 29 & 4 &$-132$ & Trunc % early 
          \\

fp64 & fp64 & H100/H200/B200/RTX PRO & - & - & 1 & - & all 4  %early
\\
          & & A100 & - & - & 1 & - & all 4  %early
          \\
          \bottomrule
\end{tabular}}
\end{center}
{\footnotesize
Note: subnormal in/out supported;
output in fp16 format is also supported for fp16 input format with RNE as the final rounding.
The products remain denormalised in alignment and accummulation, reflected via $2$ integer bits.
Extra alignment bits $\neab$ are included in the fractional bits of the product alignment precision.
Accumulation output precision is a sum of extra carry bits and the product alignment bits. RTX PRO refers to RTX PRO 6000. 
%* fp8 implementation with \texttt{mma.sync.aligned.m16n8k16/32.f32/f16.f8.f8.f32/f16} where fp8 to fp16 conversion takes place and \texttt{HMMA} instructions are called.
}
\end{table*}

%----------------------

\begin{table*}
  \begin{center}
  \caption{CUDA/PTX instrunction mapping to SASS instructions}\label{tab:cf}
% Second version of table, with booktabs.
\begin{tabular}{lllr}\toprule

  Input & CUDA/PTX Instr. & GPUs & SASS Instr. \\\midrule
fp8 & \texttt{mma.sync.aligned.m16n8k32} & H100/H200/B200 & \texttt{HMMA.16816} \\
         & & RTX 6000 & \texttt{QMMA.16832}
\\
         & & L40S/Ada RTX 1000 & \texttt{QMMA.16816}  \\

 & \texttt{wgmma.mma\_async.sync.m64nNk32} & H100/H200  & \texttt{QGMMA.16832} \\

 & \texttt{tcgen05.mma.cta\_group} & B200  & \texttt{UTCQMMA} \\\hline

fp16/bf16 & {WMMA API} & H100/H200/B200/RTX PRO & \texttt{HMMA.16816}  \\
                  & & L40S/Ada RTX 1000 & \texttt{HMMA.1688}\\
                  & & A100/A2/A30/A40 & \texttt{HMMA.1688} \\
                  & & V100 & \texttt{HMMA.844}\\
 & \texttt{tcgen05.mma.cta\_group.kind=f16} & B200  & \texttt{UTCHMMA} \\\hline
tf19 & WMMA API & H100/H200/B200 & \texttt{HMMA.1684}  \\
          & & L40S/Ada RTX 1000 & \texttt{HMMA.1684} \\
          & & A100/A2/A30/A40 & \texttt{HMMA.1684}  \\
 & \texttt{mma.sync.aligned.m16n8k4/8} & H100/H200/B200/RTX PRO & \texttt{HMMA.1684/8}  \\
          & & L40S/Ada RTX 1000 & \texttt{HMMA.1684} \\
          & & A100/A2/A30 & \texttt{HMMA.1684}  \\
 & \texttt{tcgen05.mma.cta\_group.kind=tf32} & B200  & \texttt{UTCHMMA} \\\hline
fp64 & WMMA API & H100/H200/B200/RTX PRO & \texttt{DMMA.884}  \\
          & & A100 & \texttt{DMMA.884} \\\bottomrule
\end{tabular}
\end{center}
{\footnotesize
WMMA API for tf19 maps to HMMA.1684 for GPUs that support tf19 as input, but H100/H200/B200 and RTX PRO (refers to RTX PRO 6000) also support HMMA.1688 via \texttt{mma.sync.aligned.m16n8k8}. 
The \texttt{wgmma.mma} PTX instruction is specific to the Hopper architecture, whereas the \texttt{tcgen05.mma} PTX instruction is specific to Blackwell data-center GPUs.
}
\end{table*}

\subsection{Comparison of Software Models Against GPU Results: Sufficiency of Ensemble Testing and Minimum Amount of Testing Vectors}

For increased assurance that tensor core models are bit-equivalent to GPU tensor core and no additional feature remains undetected, we have
compared the tensor core models against the GPU results over 
ensembles containing vectors of following cases.
\begin{enumerate}
\item Sampling $a$, $b$, and $c$ from independent and identically distributed (i.i.d) uniform distribution with upper and lower bounds denoted as $a_{\mathrm{min}},~a_{\mathrm{max}},~b_{\mathrm{min}},~b_{\mathrm{max}},~c_{\mathrm{min}},~c_{\mathrm{max}}$, respectively, for each term with ensemble size denoted as $N_{\mathrm{ens}}$. 
  While \(c\) is a scalar, \(a\) and \(b\) are vectors of length \(k\), where $k$ is the inner dimension supported by a single MMA instruction. 
  Note that $k$ is not necessarily equal to $\NFMA$, because some of the PTX instructions support larger inner dimension than that supported by hardware tensor cores. 
  For instance $k=16$ for Ampere and Ada in fp16/bf16 input format but $\NFMA=8$.
 \begin{itemize}
    \item In input formats {bf16} and {tf19}, we generate ensemble via Algorithm~\ref{alg:ensemble_genertion}, where $k=16$
      for bf16, and $8$ for tf19, with following settings.
      \begin{enumerate}
    \item $N_{\mathrm{ens}}=10^7$,  $a_{\mathrm{min}}=b_{\mathrm{min}}=c_{\mathrm{min}}=-2^{15}$, $a_{\mathrm{max}}=b_{\mathrm{max}}=c_{\mathrm{max}}=2^{15}$
    %% extended range
    \item $N_{\mathrm{ens}}=10^7$,  $a_{\mathrm{min}}=b_{\mathrm{min}}=-2^{50}$, $a_{\mathrm{max}}=b_{\mathrm{max}}=2^{50}$, $c_{\mathrm{min}}=-2^{100},~c_{\mathrm{max}}=2^{100}$
    %% subnormal region
    \item $N_{\mathrm{ens}}=10^7$, $a_{\mathrm{min}}=b_{\mathrm{min}}=-2^{-70}$, $a_{\mathrm{max}}=b_{\mathrm{max}}=2^{-70},~c_{\mathrm{min}}=-2^{-126}$, $c_{\mathrm{max}}=2^{-126}$
    \item $N_{\mathrm{ens}}=10^7$,    $a_{\mathrm{min}}=b_{\mathrm{min}}=-2^{-70}$, $a_{\mathrm{max}}=b_{\mathrm{max}}=2^{-70}$, $c=0$
      \end{enumerate}
    First ensemble tests the basic working accuracy within narrow range i.e. $\pm 2^{15}$ of the mentioned format, and thereafter, models are tested within extended range, but avoiding overflows both in input and output.
    In the last two cases, subnormal region is explored where the product exponents are intentionally made smaller than $e_{\mathrm{min}}^{\mathrm{lim}}$ with $c$ either in subnormal region or set to $0$.

%%
% fp16/fp8-e5m2
%%

  \item In input formats {fp16} and {fp8-e5m2}, ensembles are generated with $k=16$ for fp16 and $k=32$ for fp8-e5m2, for all GPUs that support these formats.
    \begin{enumerate}
    \item $N_{\mathrm{ens}}=10^7$, $a_{\mathrm{min}}=b_{\mathrm{min}}=c_{\mathrm{min}}=-2^{7}$,
      $a_{\mathrm{max}}=b_{\mathrm{max}}=c_{\mathrm{max}}=2^{7}$
    \item $N_{\mathrm{ens}}=10^7$, $a_{\mathrm{min}}=b_{\mathrm{min}}=-2^{15},~a_{\mathrm{max}}=b_{\mathrm{max}}=2^{15}$,
      $c_{\mathrm{min}}=-2^{100}$, $c_{\mathrm{max}}=2^{100}$
    \item $N_{\mathrm{ens}}=10^7$, $a_{\mathrm{min}}=b_{\mathrm{min}}=-2^{-15}$, $a_{\mathrm{max}}=b_{\mathrm{max}}=2^{-15}$, $c_{\mathrm{min}}=-2^{-126}$, $c_{\mathrm{max}}=2^{-126}$
      \end{enumerate}
    Similar to the bf16 and tf19 case, these three cases test fp16 and fp8-e5m2 in normal and subnormal regions.

  \item In fp8-e4m3 input format, for which $L=32$, we have
    \begin{enumerate}
    \item $N_{\mathrm{ens}}=10^7$, $a_{\mathrm{min}}=b_{\mathrm{min}}=-2^{7}$,
      $a_{\mathrm{max}}=b_{\mathrm{max}}=2^{7},~c_{\mathrm{min}}=-2^{100}$, $c_{\mathrm{max}}=2^{100}$
    \item $N_{\mathrm{ens}}=10^7$, $a_{\mathrm{min}}=b_{\mathrm{min}}=-2^{-6}$,
      $a_{\mathrm{max}}=b_{\mathrm{max}}=2^{-6}$, $c_{\mathrm{min}}=-2^{-126}$, $c_{\mathrm{max}}=2^{-126}$
 \end{enumerate}
 %% H100/H200/B200
    \item In {bf16} format, we generate an extended ensemble with $k=16$ for Ampere, Hopper, and Blackwell GPUs
    with
    $N_{\mathrm{ens}}=10^8$, $a_{\mathrm{min}}=b_{\mathrm{min}}=-2^{63}$, $a_{\mathrm{max}}=b_{\mathrm{max}}=2^{63}$, $c_{\mathrm{min}}=-2^{122}$, $c_{\mathrm{max}}=2^{122}$. This setup enables testing of the model across all
possible output categories, including Inf, NaN, normal, and subnormal values.

\end{itemize}

   \item $N_{\mathrm{ens}}=10^5$ samples drawn via Algorithm~\ref{alg:ensemble_genertion} from a normal distribution in all input formats.
   \item Edge Cases
    \begin{itemize}
    \item Test vectors to evaluate behaviour for $\pm\texttt{Inf}$ and \texttt{NaN} inputs.
    \item Scenarios where \( c = 0 \), or where some or all partial products are zero.
    \item Extreme dynamic range cases, e.g., \( c = \pm 2^{127} \) with products \( -2^{-126} \leq p_i \le 2^{-126} \), and vice versa.
    \item Subnormal cases, with \( c = 0 \) or a subnormal value in fp32 format and \( p_i \in [-2^{-126}, 2^{-126}] \).
    \item Cases where partial products exceed \( 2^{128} \), but the final sum is less than $2^{128}$.
    \item Mixed input configurations involving zero, normal, subnormal, and special values.
    \item Choose $a,~b$, and $c$ such dot product generates the maximum number of carries during accumulation (see~\cite{khmi25}).
\end{itemize}
  \end{enumerate}

%% ===================
\begin{algorithm}
\caption{Ensemble generation psuedo-code for randomized testing of tensor core models}
\label{alg:ensemble_genertion}

\begin{algorithmic}[1]

\State Choose $k$ as per input format and GPU model
\State mt19937 rng(0) \quad $\%~$Mersenne Twister with seed 0
\State choose $N_{\mathrm{ens}}$, $a_{\mathrm{min}}$, $a_{\mathrm{max}}$, $b_{\mathrm{min}}$, $b_{\mathrm{max}}$, $c_{\mathrm{min}}$, $c_{\mathrm{max}}$ 
\For{$i=1:N_{\mathrm{ens}}$}
\State $a_\ell\overset{\mathrm{i.i.d.}}{\sim}\mathcal{U}\left(a_{\mathrm{min}},a_{\mathrm{max}}\right)$, $\ell=1,\dots,k$
\State $b_\ell\overset{\mathrm{i.i.d.}}{\sim}\mathcal{U}\left(b_{\mathrm{min}},b_{\mathrm{max}}\right)$, $\ell=1,\dots,k$
\State $c\overset{\mathrm{i.i.d.}}{\sim}\mathcal{U}\left(c_{\mathrm{min}},c_{\mathrm{max}}\right)$
\EndFor

\end{algorithmic}
\end{algorithm}

%==============================

For the hardware execution, \texttt{WMMA} API, \texttt{mma.sync}, \texttt{wgmma.mma\_async}, and \texttt{tcgen05.mma} instructions were invoked in CUDA.
To further verify that tensor cores were active, the \texttt{cuobjdump} tool was used to inspect the compiled binary and confirm the presence of \texttt{HMMA}, \texttt{QMMA}, \texttt{QGMMA}, \texttt{UTCHMMA}, \texttt{UTCQMMA}, \texttt{DMMA} instructions.
Finally, we note that \texttt{DMMA} operations are not modelled or tested via randomized testing, nor are they included in the model package, since, as reported by Fasi et al.~\cite{fhmp21}, they behave as sequential FMAs and are fully IEEE compliant.
Therefore, one can directly rely on MATLAB's built-in \texttt{fma} command to emulate the behaviour of \texttt{DMMA}.

For Ada GPUs, CUDA toolkit 12.9 was used, while for all other GPUs, we used CUDA toolkit 12.8, while the GCC version for Ada GPUs and B200 was 14.2.0, and 13.1.0, respectively, while for remaining GPUs, it was 11.5.0.

\subsubsection{Evolution of Software Models Across Iterations of Model Refinement}
Here we describe how the models presented in this paper evolved over the iterations of Algorithm~\ref{alg:refine-model}.

In the first iteration, using GNFT, we determined the FMA sizes, extra alignment and carry bits, alignment rounding mode, and final rounding modes (truncation in fp32, and RNE for fp16 output format). 
However, the GNFT-only model failed for certain test cases in the ensemble, which subsequently led to the identification of the denormalised product feature. 
Consequently, the GNFT was extended with dedicated test vectors capable of detecting this behaviour. 
This feature also revealed that products are computed as the product of significands multiplied by two raised to the sum of exponents, thereby allowing intermediate products to exceed $2^{128}$ without overflow, provided that the final accumulated result remains representable in the output format. 

After incorporating the necessary updates into the software models, the second iteration showed that the V100 model satisfied the entire test ensemble, as it only supports fp16 input formats. 
However, all other models failed within regions of the ensemble involving subnormal \(a\), \(b\), and \(c\) values in bf16 and tf19 formats (prompted by a GitHub issue report from a commonity member, username \emph{dfyz}).
Detailed analysis revealed that the minimum exponent permitted during product alignment is $-132$ for Ampere and Ada, and $-133$ for Hopper and Blackwell in both formats, with \(c=0\) treated as a special case, i.e., ignored rather than interpreted as a subnormal value. 
This observation prompted the construction of dedicated tests to identify exponent limitations, which were subsequently incorporated into the GNFT.
In the third iteration, the GPU and \textsc{MATLAB} model outputs matched across the entire test ensemble, at the bit level, across all supported input and output precision formats across 11 GPUs for 8-bit and higher precision formats (Table~\ref{tab:nf}).

Moreover, the emulated models were also updated to match the GPU outputs in exceptional cases (\texttt{NaN}, \texttt{Inf}, or \texttt{-Inf}) via expressions such as $\infty \pm \infty$, $\pm(\texttt{Inf} \times \texttt{Inf})$, and $\texttt{NaN} \pm \texttt{Inf}$, both as direct inputs and as overflow scenarios where finite accumulations lead to $\pm \texttt{Inf}$ in the supported output formats.
Consistent with GPU behavior, \texttt{NaN} takes precedence
whenever it appears.

%================================
\subsubsection{Sufficiency of the Constructed Ensemble}

The sufficiency of the randomized testing strategy is supported by the structural simplicity of the proposed models. 
In particular, the truncation of aligned significands beyond $(23+\neab)-$fractional bits eliminates many complex corner cases, such as partially or completely out-of-range shifted significands, thereby significantly reducing the effective search space. 
Consequently, broad randomized sampling, combined with targeted stress testing, provides high coverage of relevant numerical behaviours.

The constructed ensemble consists of complementary randomized and targeted tests designed to exercise the models across normal, subnormal, and special-value operating regions. The ensemble includes evaluations over broad dynamic ranges, targeted subnormal-region analysis for different tensor core formats, and extended stress testing capable of triggering all output categories, including normal, subnormal, Inf, and NaN outputs. In addition, edge cases were incorporated to evaluate behaviours involving zero-valued accumulators or products, exponent alignment limitations, extreme dynamic ranges, denormalised products, and mixed combinations of normal, subnormal, and special values.

Together, the combination of randomized sampling, targeted edge-case evaluation, and iterative refinement through Algorithm~1 provides strong empirical evidence that the proposed models accurately capture the numerical behaviour of the hardware. 
While exhaustive validation over the complete input space is infeasible, the constructed ensemble provides a high degree of confidence in bit-level agreement between the software models and GPU implementations.
The extended ensemble testing over 100 million input samples was performed and the absence of further mismatches throughout these tests therefore provides strong evidence that the resulting software models are already highly accurate.

Randomized testing can also be replaced by a minimum set of test vectors for any tensor core architecture, provided that the assumed feature space is a superset of all features present in a given architecture.
However, this is not the case in the current scenario, as the initially assumed feature space (in the GNFT and prior studies of author authors) did not include denormalized products, minimum exponent limitations, and the special case of \(c=0\) instead of being considered as a subnormal case.

In the future, we will explore different strategies for ISSM; we plan to port Schryer's~\cite{schr81} strategy from testing scalar operations to dot products and explore covering signs, exponents and significands via separate sampling.
We also plan deplou the verification in parallel across hundreds of cores to cover larger portion of the input space.

%-------------------------
\subsection{Example MATLAB Code for Using the Models}

This section introduces the \texttt{MATLAB Tensor Core v0.5}.
The toolbox was developed on MATLAB R2026a and depends on CPFloat~\cite{fami23} for rounding the inputs, that are by default in binary64, to low-precision formats that MATLAB does not natively support.
Here we discuss the user interface and the overall structure of the code.

The toolbox is comprised of three layers:
\begin{itemize}
\item \texttt{Generic\_BFMA\_TC.m}: provides a generalised tensor core model which can be set up with various features~\cite{khmi25}.
\item \texttt{GEMM.m}: accepts $\alpha$, $A$, $B$, $\beta$, and $C$, floating-point input and output formats, and the settings for the \texttt{Generic\_BFMA\_TC.m}, and approximates the GEMM $\alpha \times A \times B + \beta \times C$ using the tensor core model.
\item A set of GPU tensor core models, such as \texttt{B200TC.m}, which instantiate the model parameters and call \texttt{GEMM.m}.
\end{itemize}

The v0.5 of the toolbox implements a recursive algorithm in \texttt{GEMM.m}, which is equivalent to recursively using a single tensor core to compute each inner product in the GEMM. It does not attempt to match the results of any CUDA GEMM algorithm, of which there are possibly many and they are chosen dynamically based on the input matrix shapes and numerical formats.
The following shows an example that calls the B200 tensor core model with an fp8 format as an input and fp32 as the output for multiplying two $4\times 4$ matrices.

\begin{tcolorbox}[colback=gray!10, colframe=gray!30,
top=0pt, bottom=0pt, left=3pt, right=0pt,   % remove internal margins
    boxsep=0pt,]
\begin{lstlisting}
inopts.format = 'fp8-e4m3', outopts.format = 'binary32';
A = cpfloat(randn(4), inopts), B = cpfloat(randn(4), inopts);
C = cpfloat(randn(4), outopts), alpha = 1, beta = 1;
B200TC(alpha, A, B, beta, C, inopts.format, outopts.format)
>> ans =
    0.1484   -0.6631   -0.1836   -1.3271
    0.8232    1.6418    0.4805    3.0227
    3.6592   -0.1250    1.4902    2.2637
    3.7432   -3.4275    0.2031    0.1663
\end{lstlisting}
\end{tcolorbox}

Apart from calling the mentioned GPUs TC functions, the user can call a custom model TC function with a set of parameters; an example is provided below.

\begin{tcolorbox}[colback=gray!10, colframe=gray!30,
top=0pt, bottom=0pt, left=2pt, right=0pt,   % remove internal margins
    boxsep=0pt,]
\begin{lstlisting}
params.neab = 3;       % Extra alignment bits.
params.fma = 32;       % Block FMA size.
params.frmode = 'rne'; % Final rounding mode.
CustomTC(alpha, A, B, beta, C, in_format, out_format, params)
\end{lstlisting}
\end{tcolorbox}

% A direct example of CustomTC model is provided in the form of \texttt{B200TCRN} which uses B200 features but with \texttt{params.frmode} set to \texttt{rne}.
% There is also a field \texttt{stkbitenabled} in \texttt{params}, i.e., \texttt{params.stkbitenabled}, which, when the significands of product terms are aligned, appends an extra sticky bit beyond the extra alignment bits ($\neab$).
% By default, this is set to 0.
% In addition, the field \texttt{params.inter\_pattern}, set to zero by default, is included due to the H100/H200/B200 tensor core behavior with the fp8 input format (Section~\ref{sec:h100})
% when it is accessed through \texttt{mma.sync}.
% If this field is set to $1$, the custom model will compute the inner product of $2\NFMA$-element vectors by interleaving the elements, as per Figure~\ref{fig_H100}c, with the input $c$ added to the product at the end, using a RNE rounding mode.
% If this filed is turned off, it implements \texttt{QGMMA} in Hopper architecture and \texttt{UTCQMMA} in Blackwell for fp8 input format.

The \texttt{GEMM.m} file executes the parallelized version of matrix multiplication if the \texttt{Parallel Computing Toolbox}, introduced in Matlab R2013b, is installed and the machine supports multicore processing.
This ensures that computations are efficiently distributed across available CPU cores, accelerating large-scale matrix operations.
If either is not supported, a serialized version is executed.
Lastly, the proposed toolbox is compatible with Octave and can also be accessed from Python using either Oct2Py library or the Matlab engine.
In the future, we will port the back-end of the models to C and interface to MATLAB via the mex interface, to improve performance of computing with the tensor core models.
%------------------------------------------------------
%- Comparison and Success Rate ------------------------
%------------------------------------------------------
\subsection{Comparison of our Models with Prior Models}

\subsubsection{Numerical Features}

A comparison of the numerical features identified by different studies is presented in Table~\ref{tab:comp_feat_table}, including prior works such as \cite{hibr19,llfs24,vlpg25,khmi25,fhmp21}. Hickmann et al.~\cite{hibr19} investigated only the V100 GPU and correctly identified most numerical features. However, they concluded that the $c$ term is added late to the product sum using RNE, whereas our results indicate otherwise. Furthermore, the preservation of denormalized products was not reported.

Similarly, Fasi et al.~\cite{fhmp21} examined the tensor cores of both V100 and A100 GPUs. While their reported features are largely consistent with our findings, they did not identify the preservation of denormalized products, nor the exponent limitation present in the bf16 format on A100. Li et al.~\cite{llfs24} extended the analysis to H100 GPUs in addition to V100 and A100, but did not investigate the fp8 input format. Moreover, they did not report either the denormalized product behavior or the exponent limitation feature. Their characterization of $\NFMA$ and $\neab$ for H100 with bf16 and fp16 inputs was also inconclusive, reporting only bounds, i.e., $\neab \geq 2$ and $\NFMA \geq 16$.

The GNFT-only studies likewise did not identify the denormalized-product and exponent-limitation features and did not consider fp8 input formats. In contrast, the proposed framework combines targeted feature testing with randomized numerical testing, enabling the discovery of previously unreported behaviors and hidden numerical characteristics across multiple GPU architectures and input precisions. As shown in Table~\ref{tab:comp_feat_table}, the proposed approach provides the most comprehensive coverage of tensor-core numerical features among the compared works.

\subsubsection{Ensemble Based Failure Rate}

Here we compare the refined model, the GNFT-only model, and prior models against GPU outputs in terms of ensemble failure rate. 
While no publicly available software implementations exist for any of the earlier studies, we use the proposed \textsc{Matlab} model to emulate the models in~\cite{hibr19,llfs24,fhmp21}, since these papers provide comprehensive descriptions of the numerical features required for accurate emulation.
To simulate the model of~\cite{llfs24}, the following assumptions are made:
\begin{itemize}
    \item $\NFMA = 1$ for V100, since~\cite{llfs24} reports that the FMA size is not detectable.
    
    \item Since the authors report $\neab \geq 2$ and $\NFMA \geq 16$ for H100, the minimum reported values are assumed, i.e., $\neab = 2$ and $\NFMA = 16$.
    
    \item Products are assumed to be normalised, and the minimum exponent limitation for alignment is set to $-\infty$, as these features were not detected.
\end{itemize}
The V100 model estimated in~\cite{hibr19} is simulated as reported, where $c$ is added afterwards via an RNE operation with the product normalised. 
For the GNFT-only model~\cite{khmi25}, the FMA size is correctly detected; however, product denormalisation and exponent-limitation behaviour are not incorporated.

The ensemble failure rate for an ensemble size of $10^5$ is largest for the V100 model of~\cite{hibr19}, while the second-highest failure rate is observed for the model of~\cite{llfs24}, as reported in Table~\ref{tab:comp_table}.
Although the simulated model of~\cite{llfs24} yields a failure rate similar to that of the GNFT-only model, the failure rate for H100 may in practice be even larger than that reported in Table~\ref{tab:comp_table} if simulated with $\neab > 2$ and $\NFMA > 16$.
Such settings would further deviate from the true H100 hardware behaviour.
{This is the reason for $\ge$ symbol with H100 failure rate in~\cite{llfs24} column in Table~\ref{tab:comp_table}.} 
The proposed refined model achieves zero failure rate across all illustrated test cases, while the GNFT-only model remains the closest approximation among the compared methods.

It should also be noted that, depending on how the ensemble is generated, the GNFT-only and~\cite{llfs24} models may fail to trigger certain hardware features, such as exponent limitation. Consequently, these models may exhibit zero failure rate regardless of ensemble size. An example of this behaviour is shown in the final bf16 input-format case in Table~\ref{tab:comp_table}.

\begin{table*}
  \begin{center}
  \caption{Comparison of numerical features of tensor cores on NVIDIA GPUs reported by prior works and the present work. }\label{tab:comp_feat_table}
% Second version of table, with booktabs.
\begin{tabular}{llcccccc}\toprule

 GPU & Feature  & \cite{hibr19} & \cite{fhmp21} & \cite{llfs24} & \cite{vlpg25} & \cite{khmi25} & Proposed \\\midrule

% V100
 V100 & $\NFMA$ & \ding{51}  & \ding{51} & \ding{55} & \ding{51} & \ding{51} & \ding{51} \\
 & $\neab$ & \ding{51}  & \ding{51} & \ding{51} & \ding{51} & \ding{51} & \ding{51}\\
 & denom. prod. & NR  & NR & NR & NR & NR & \ding{51}\\
 & $c$ is added late/early & \ding{55}  & \ding{51} & \ding{51} & \ding{51} & \ding{51} & \ding{51}\\\hline
% A100
 A100/A30/A40/A2 & $\NFMA$ & -  & \ding{51} & \ding{51} & \ding{51} & \ding{51} & \ding{51} \\
 & $\neab$ & -  & \ding{51} & \ding{51} & \ding{51} & \ding{51} & \ding{51}\\
 & denom. prod. & -  & NR & NR & NR & NR & \ding{51}\\
 & exp. limit. ($e^{\mathrm{align}}_{\mathrm{min}}$) & -  & NR & NR & NR & NR & \ding{51}\\
 & $c=0$ special/subnormal case & -  & NR & NR & NR & NR & \ding{51}\\\hline

%% Ada Lovelace
 Ada (RTX-1000/L40S) & $\NFMA+\neab$ (fp16/bf16/tf19) & -  &- & - & - & \ding{51} & \ding{51} \\

 & $\NFMA+\neab$ (fp8) & 
-  & - & - & - & - & \ding{51}\\

 & denom. prod. & 
-  & - & - & - & NR & \ding{51}\\

 & exp. limit. ($e^{\mathrm{align}}_{\mathrm{min}}$) & -  & - & - & - & NR & \ding{51}\\

 & $c=0$ special/subnormal case & 
-  & - & - & - & NR & \ding{51}\\\hline

%% Hopper
 H100/H200 & $\NFMA+\neab$ (fp16/bf16/tf19) & -  &- & \ding{55}* & - & - & \ding{51} \\

 & $\NFMA+\neab$ (fp8) & 
-  & - & - & - & - & \ding{51}\\

 & denom. prod. & 
-  & - & NR & -  & - & \ding{51}\\

 & exp. limit. ($e^{\mathrm{align}}_{\mathrm{min}}$) & -  & - & NR & -  & - & \ding{51}\\

 & $c=0$ special/subnormal case & 
-  & - & NR & -  & - & \ding{51}\\\hline

%% Blackwell
 B200/RTX PRO 6000 & $\NFMA+\neab$ (fp16/bf16/tf19) & -  &- & - & - & - & \ding{51} \\

 & $\NFMA+\neab$ (fp8) & 
-  & - & - & - & - & \ding{51}\\

 & denom. prod. & 
-  & - & - & - & - & \ding{51}\\

 & exp. limit. ($e^{\mathrm{align}}_{\mathrm{min}}$) & -  & - & - & - & - & \ding{51}\\

 & $c=0$ special/subnormal case & 
-  & - & - & - & - & \ding{51}

\\\bottomrule
\end{tabular}
\end{center}
{\footnotesize
Cells marked - mean that GPU or the input format is not considered, \ding{55}: feature tested but reported conclusion is inaccurate,~NR: stands for not reported,~ \ding{55}*: feature detected partially correct or reported with ambiguity}.
\end{table*}

\begin{table*}
  \begin{center}
  \caption{Ensemble-based mismatch rates of proposed and prior tensor core models compared against GPU outputs.}\label{tab:comp_table}
% Second version of table, with booktabs.
\begin{tabular}{lrllr}\toprule

 Ensemble setting  & \cite{hibr19} & \cite{llfs24} & GNFT~\cite{khmi25} & Proposed \\\midrule

 (fp16): $a$, $b$, $c\in[-2^7,2^7]$ & V100: $55.4\%$ & $45\%$  & $22.5\%$ & $0\%$ \\

 & A100: NA & $15.5\%$ & $15.5\%$  & $0\%$\\

 & H100: NA & $13\%$ & $13\%$  & $0\%$ \\\hline

 (bf16): $a,b,c\in[-2^{15},2^{15}]$ & A100: NA & $9.6\%$ & $9.6\%$  & $0\%$\\

 & H100: NA & $\ge 9.2\%$ & $9.2\%$  & $0\%$ \\
 % subnormal case

$c=0$, $a$, $b\in[-2^{-70},2^{-70}]$ & A100: NA & $0.006\%$ & $0.006\%$  & $0\%$\\

 & H100: NA & $\ge 0.002\%$ & $0.002\%$  & $0\%$ \\
  % subnormal case
 $c=[-2^{-126},2^{-126}]$, $a$, $b\in[-2^{-65},2^{-65}]$ & A100: NA & $0\%$ & $0\%$  & $0\%$\\

 & H100: NA & $\ge 0\%$ & $0\%$  & $0\%$ 
\\\bottomrule
\end{tabular}
\end{center}
{\footnotesize
The ensemble for comparison is created with a seed of $0$ via mt19937 by sampling $10^5$ samples from a uniform distribution}.
\end{table*}

\section{Multi-Word Algorithms for Emulating High-Precision GEMM on Tensor Cores Models}

\begin{figure*}
  \begin{center}
    \footnotesize
    \begin{tikzpicture}
      \begin{groupplot}[
        group style={
          group size=3 by 3,
          vertical sep=1.2cm
        },
        ymode=log,
        xmode=log,
        width=1.9in,
        height=1.6in,
        grid=major,
        ymax = 10^(0),
        ymin = 10^(-9),
        every axis plot/.append style={very thick, mark repeat=3}
        ]

        \nextgroupplot[
        ylabel={$\frac{\norminf{\widehat{C}-C}}{\norminf{A}\norminf{B}}$},
        align=left,
        title={Single word (fp8-e5m2)},
        xlabel = {$n$}
        ]

        \addplot[color=Magenta!70, mark=square] table [x=n, y=error] {data/matmul_test_fp8-e5m2_binary32_words_1_model_h100.dat};
        \addplot[color=blue!70, mark=x] table [x=n, y=error] {data/matmul_test_fp8-e5m2_binary32_words_1_model_b200.dat};
        \addplot[color=black!70, dashdotted] table [x=n, y=error] {data/matmul_test_fp8-e5m2_binary32_words_1_model_l40s.dat};
        \addplot[color=red!70, mark=o] table [x=n, y=error] {data/matmul_test_fp8-e5m2_binary32_words_1_model_b200rn.dat};

        \nextgroupplot[
        align=left,
        title={Quad-word (fp8-e5m2)},
        xlabel = {$n$}
        ]

        \addplot[color=Magenta!70, mark=square] table [x=n, y=error] {data/matmul_test_fp8-e5m2_binary32_words_4_model_h100.dat};
        \addplot[color=blue!70, mark=x] table [x=n, y=error] {data/matmul_test_fp8-e5m2_binary32_words_4_model_b200.dat};
        \addplot[color=black!70, dashdotted] table [x=n, y=error] {data/matmul_test_fp8-e5m2_binary32_words_4_model_l40s.dat};
        \addplot[color=red!70, mark=o] table [x=n, y=error] {data/matmul_test_fp8-e5m2_binary32_words_4_model_b200rn.dat};

        \nextgroupplot[
        align=left,
        title={6-word (fp8-e5m2)},
        xlabel = {$n$}
        ]

        \addplot[color=Magenta!70, mark=square] table [x=n, y=error] {data/matmul_test_fp8-e5m2_binary32_words_6_model_h100.dat};
        \addplot[color=blue!70, mark=x] table [x=n, y=error] {data/matmul_test_fp8-e5m2_binary32_words_6_model_b200.dat};
        \addplot[color=black!70, dashdotted] table [x=n, y=error] {data/matmul_test_fp8-e5m2_binary32_words_6_model_l40s.dat};
        \addplot[color=red!70, mark=o] table [x=n, y=error] {data/matmul_test_fp8-e5m2_binary32_words_6_model_b200rn.dat};

      \end{groupplot}
    \end{tikzpicture}

    \begin{tikzpicture}[trim axis left, trim axis right]
      \begin{axis}[
        title = {},
        legend columns=5,
        scale only axis,
        width=1mm,
        hide axis,
        /tikz/every even column/.append style={column sep=0.4cm},
        legend style={at={(0,0)},anchor=center,draw=none,
          legend cell align={left},cells={line width=0.75pt}},
        legend image post style={sharp plot},
        legend cell align={left},
        every axis plot/.append style={thick}
        ]
        \addplot [black!70, dashdotted] (0,0);
        \addplot [color=Magenta!70, mark=square] (0,0);
        \addplot [color=blue!70, mark=x] (0,0);
        \addplot [red!70, mark=o] (0,0);
        \legend{L40S, H100, B200, B200 RN};
      \end{axis}
    \end{tikzpicture}
  \end{center}
  \caption{Multi-word arithmetic experiment presented by Mary and Mikaitis~\cite[Sec.~5]{mami25} on the simulation of various tensor cores. We have reproduced the experiment on four different tensor core generations modelled in MATLAB. Relative norm-wise errors of matrix multiplication, compared with a default MATLAB binary64 multiplication, are shown. The input matrices to the GEMM are $A \in R^{10\times n}$ and $B \in R^{n \times 10}$. These matrices are multiplied with a multi-word algorithm \cite[Sec.~4]{mami25} by splitting them into several fp8-e5m2 words.}
   \label{fig:multi-word0}
\end{figure*}
\begin{figure*}
  \begin{center}
    \footnotesize
    \begin{tikzpicture}
      \begin{groupplot}[
        group style={
          group size=3 by 3,
          vertical sep=1.2cm
        },
        ymode=log,
        xmode=log,
        width=1.9in,
        height=1.6in,
        grid=major,
        ymax = 10^(0),
        ymin = 10^(-9),
        every axis plot/.append style={very thick, mark repeat=3}
        ]

        \nextgroupplot[
        ylabel={$\frac{\norminf{\widehat{C}-C}}{\norminf{A}\norminf{B}}$},
        align=left,
        title={Single word (binary16)},
        xlabel = {$n$}
        ]

        \addplot[color=black!70, mark=diamond] table [x=n, y=error] {data/matmul_test_binary16_binary32_words_1_model_v100.dat};
        \addplot[color=black!70, dashed] table [x=n, y=error] {data/matmul_test_binary16_binary32_words_1_model_a100.dat};
        \addplot[color=blue!70, mark=x] table [x=n, y=error] {data/matmul_test_binary16_binary32_words_1_model_b200.dat};
        \addplot[color=red!70, mark=o] table [x=n, y=error] {data/matmul_test_binary16_binary32_words_1_model_b200rn.dat};

        \nextgroupplot[
        align=left,
        title={Double word (binary16)},
        xlabel = {$n$}
        ]

        \addplot[color=black!70, mark=diamond] table [x=n, y=error] {data/matmul_test_binary16_binary32_words_2_model_v100.dat};
        \addplot[color=black!70, dashed] table [x=n, y=error] {data/matmul_test_binary16_binary32_words_2_model_a100.dat};
        \addplot[color=blue!70, mark=x] table [x=n, y=error] {data/matmul_test_binary16_binary32_words_2_model_b200.dat};
        \addplot[color=red!70, mark=o] table [x=n, y=error] {data/matmul_test_binary16_binary32_words_2_model_b200rn.dat};

        \nextgroupplot[
        align=left,
        title={Triple word (binary16)},
        xlabel = {$n$}
        ]

        \addplot[color=black!70, mark=diamond] table [x=n, y=error] {data/matmul_test_binary16_binary32_words_3_model_v100.dat};
        \addplot[color=black!70, dashed] table [x=n, y=error] {data/matmul_test_binary16_binary32_words_3_model_a100.dat};
        \addplot[color=blue!70, mark=x] table [x=n, y=error] {data/matmul_test_binary16_binary32_words_3_model_b200.dat};
        \addplot[color=red!70, mark=o] table [x=n, y=error] {data/matmul_test_binary16_binary32_words_3_model_b200rn.dat};

      \end{groupplot}
    \end{tikzpicture}

    \begin{tikzpicture}[trim axis left, trim axis right]
      \begin{axis}[
        title = {},
        legend columns=5,
        scale only axis,
        width=1mm,
        hide axis,
        /tikz/every even column/.append style={column sep=0.4cm},
        legend style={at={(0,0)},anchor=center,draw=none,
          legend cell align={left},cells={line width=0.75pt}},
        legend image post style={sharp plot},
        legend cell align={left},
        every axis plot/.append style={thick}
        ]
        \addplot [black!70, mark=diamond] (0,0);
        \addplot [black!70, dashed] (0,0);
        \addplot [color=blue!70, mark=x] (0,0);
        \addplot [red!70, mark=o] (0,0);
        \legend{V100, {L40S, A100}, {H100, H200, B200}, B200 RN};
      \end{axis}
    \end{tikzpicture}
  \end{center}
  \caption{Multi-word arithmetic experiment presented by Mary and Mikaitis~\cite[Sec.~5]{mami25} on the simulation of various tensor cores. Relative norm-wise errors of matrix multiplication, compared with a default MATLAB binary64 multiplication, are shown. The input matrices to the GEMM are $A \in R^{10\times n}$ and $B \in R^{n \times 10}$. These matrices are multiplied with a multi-word algorithm \cite[Sec.~4]{mami25} by splitting them into several fp16 words.}
   \label{fig:multi-word1}
\end{figure*}
\begin{figure*}
  \begin{center}
    \footnotesize
    \begin{tikzpicture}
      \begin{groupplot}[
        group style={
          group size=3 by 3,
          vertical sep=1.2cm
        },
        ymode=log,
        xmode=log,
        width=1.9in,
        height=1.6in,
        grid=major,
        ymax = 10^(0),
        ymin = 10^(-9),
        every axis plot/.append style={very thick, mark repeat=3}
        ]

        \nextgroupplot[
        ylabel={$\frac{\norminf{\widehat{C}-C}}{\norminf{A}\norminf{B}}$},
        align=left,
        title={Single word (bfloat16)},
        xlabel = {$n$}
        ]

        \addplot[color=black!70, dashed] table [x=n, y=error] {data/matmul_test_bfloat16_binary32_words_1_model_a100.dat};
        \addplot[color=blue!70, mark=x] table [x=n, y=error] {data/matmul_test_bfloat16_binary32_words_1_model_b200.dat};
        \addplot[color=red!70, mark=o] table [x=n, y=error] {data/matmul_test_bfloat16_binary32_words_1_model_b200rn.dat};

        \nextgroupplot[
        align=left,
        title={Double word (bfloat16)},
        xlabel = {$n$}
        ]

        \addplot[color=black!70, dashed] table [x=n, y=error] {data/matmul_test_bfloat16_binary32_words_2_model_a100.dat};
        \addplot[color=blue!70, mark=x] table [x=n, y=error] {data/matmul_test_bfloat16_binary32_words_2_model_b200.dat};
        \addplot[color=red!70, mark=o] table [x=n, y=error] {data/matmul_test_bfloat16_binary32_words_2_model_b200rn.dat};

        \nextgroupplot[
        align=left,
        title={Triple word (bfloat16)},
        xlabel = {$n$}
        ]

        \addplot[color=black!70, dashed] table [x=n, y=error] {data/matmul_test_bfloat16_binary32_words_3_model_a100.dat};
        \addplot[color=blue!70, mark=x] table [x=n, y=error] {data/matmul_test_bfloat16_binary32_words_3_model_b200.dat};
        \addplot[color=red!70, mark=o] table [x=n, y=error] {data/matmul_test_bfloat16_binary32_words_3_model_b200rn.dat};

      \end{groupplot}
    \end{tikzpicture}

    \begin{tikzpicture}[trim axis left, trim axis right]
      \begin{axis}[
        title = {},
        legend columns=5,
        scale only axis,
        width=1mm,
        hide axis,
        /tikz/every even column/.append style={column sep=0.4cm},
        legend style={at={(0,0)},anchor=center,draw=none,
          legend cell align={left},cells={line width=0.75pt}},
        legend image post style={sharp plot},
        legend cell align={left},
        every axis plot/.append style={thick}
        ]
        \addplot [black!70, dashed] (0,0);
        \addplot [color=blue!70, mark=x] (0,0);
        \addplot [red!70, mark=o] (0,0);
        \legend{ {L40S, A100}, {H100, H200, B200}, B200 RN};
      \end{axis}
    \end{tikzpicture}
  \end{center}
  \caption{Multi-word arithmetic experiment presented by Mary and Mikaitis~\cite[Sec.~5]{mami25} on the simulation of various tensor cores. Relative norm-wise errors of matrix multiplication, compared with a default MATLAB binary64 multiplication, are shown. The input matrices to the GEMM are $A \in R^{10\times n}$ and $B \in R^{n \times 10}$. These matrices are multiplied with a multi-word algorithm \cite[Sec.~4]{mami25} by splitting them into several bfloat16 words.}
   \label{fig:multi-word2}
\end{figure*}

Multi-word arithmetic is a technique of emulating high-precision matrix multiplication with low precision tensor cores.
It consists of splitting high-precision input matrices into several low-precision matrices, multiplying them with tensor cores, and adding up the products either in tensor cores or the CUDA cores.
In order to demonstrate how one may use the Matlab tensor core models, we have reproduced the experiments of Mary and Mikaitis~\cite[Sec.~5]{mami25} via several different models.
We refer the readers to \cite{mami25} for a full description of the algorithms.

Figures~\ref{fig:multi-word0}--\ref{fig:multi-word2} show the norm-wise errors, where $\lVert\cdot\rVert_{\infty}$ is the infinity norm of a matrix, on five different tensor core models (V100, A100, the H100 which also covers H200 and the B200, L40S, and a custom variant of the B200 which uses round to nearest instead of bit truncation) with the input format set to fp8, fp16 and bf16, and the output format set to fp32.
The number of words, which defines how accurate the emulation is, is shown at the top of each of the sub-figures.

The modified B200 model has rounding to nearest instead of bit truncation---this is applied when an internal accumulator is rounded to the output format, not on the alignments of significands, which are still truncated as in the standard B200 tensor core model.

Figure~\ref{fig:multi-word0} shows the differences in accuracy between L40S/H100, which have $\neab=-10$ and B200 which has $\neab=2$ in the accumulator.
There is also a significant improvement to the B200 result by enabling RN.
The L40S and H100/H200 GPUs are, as expected, the least accurate in fp8 input format because of the 13 fractional bits in the accumulator as discussed in Section~\ref{sec:l40s}.
The DeepSeek-V3 Technical Report addresses this limitation in the Hopper GPU variant H800~\cite[Sec.~3.3.2]{deep25}.

For fp16 (Figure~\ref{fig:multi-word1}), interestingly, for single-word arithmetic, all models closely match in accuracy.
For double- and triple-word arithmetic, the results of Figure~\ref{fig:multi-word1} show that the V100 tensor core provides two orders of magnitude lower error at $n=10^6$.
This may be caused by B200 tensor core model having more accurate accumulator with $\neab=2$, which is rounded to zero, making the error larger than the V100 tensor core's result.
This can occur if the V100 result is above the reference result, whilst the B200 result is closer to the reference, but below it, so that RZ pushes it further below.
We have tested this hypothesis by enabling RN in the B200, which demonstrates an improvement.
Further analysis of this behavior is out of the scope of this paper and we leave it for future work.

We expect similar differences in mixed-precision iterative refinement where GEMM operations are executed using different tensor core models~\cite{hbtd20}.
Having the tensor core models at hand, callable in MATLAB, allows researching how factors such as extra alignment bits, FMA size, and rounding modes influence the behavior and accuracy of such mixed-precision computations.

These experiments demonstrate that the proposed software models can be used by researchers to simulate and study the effects of NVIDIA GPU tensor cores with customized numerical features in diverse applications, such as in artificial intelligence and scientific computing,~and also for authentication of new test vectors for line 1 of Algorithm~\ref{alg:refine-model}.
The data and the code for producing Figures~\ref{fig:multi-word0}--\ref{fig:multi-word2} is available.\footnote{\url{https://github.com/north-numerical-computing/MATLAB-tensor-core/tree/main/experiments}}

%------------------------------------------------------------
\section{Conclusion}

We have presented the research behind the development of the \texttt{MATLAB Tensor Core v0.5} toolbox.
The toolbox contains various NVIDIA GPU tensor core models, as well as a parameterised model that can be used to instantiate custom variants of tensor cores for research purposes.
The models were verified against the hardware by large-scale randomised testing and model refinement.
The proposed toolbox includes tensor core models for NVIDIA {A2, A30, A100, Ada 1000 RTX, L40S, H100, H200, RTX PRO 6000 (Blackwell) and B200},
supporting all 8-, 16-, and 19-bit precision formats.
For binary64, NVIDIA GPU tensor cores behave as sequential chains of IEEE-compliant FMAs;
therefore, MATLAB's built-in \texttt{fma} function can be used to emulate such tensor core behaviour.
In addition to the fixed models, we provide a custom tensor core model that enables users to simulate
arbitrary configurations by adjusting the FMA size, the number of extra bits used for aligning significands,
and the rounding mode (supporting \texttt{RNE}, \texttt{RZ}, \texttt{RD}, and \texttt{RU}).
%For matrix multiplication, the toolbox utilises the MATLAB \texttt{Parallel Computing Toolbox} to create a parallel pool of workers (using the default profile). 
%This dispatches inner products within a GEMM across multiple MATLAB engines.

In the future, this toolbox will be actively maintained through regular versioned releases on GitHub. We plan to improve both the functionality and performance, by porting the back-end to C, by porting the front-end to other languages such Julia, and by adding new NVIDIA tensor core and AMD matrix engine models.
We also plan to add different GEMM algorithms and improve the validation in Algorithm~\ref{alg:refine-model} by exploring different distributions of randomised test vectors and techniques such as the ones used by Schryer~\cite{schr81}.

IEEE 754~\cite{ieee19} and OCP~\cite{rgsm23} does not standardise reduction operations and P3109~\cite{ieee25} does not require them for conformance.
We hope that a simulator of GPU matrix multiplier models will allow users to understand the differences and effects on applications of current and future matrix multiply units, and impact future standardisation efforts.

\section{Acknowledgment}
    We thank John Hodrien at University of Leeds for technical support with the Aire machine containing the NVIDIA L40S and A2, and The COSmology MAchine (COSMA) support at Durham University for providing the access to A30, V100, A100, H100, H200 and RTX PRO 6000 GPUs.
We thank Argonne National Laboratory for providing access to V100, A100, H100, and B200 GPUs.
We also thank Jack Dongarra, John Gunnels, Eduardo Basurto, and Eric Rife for arranging access to the B200 GPUs before they became available via ANL.
We are grateful to a GitHub user \emph{dfyz} (Ivan Komarov) for testing the tensor core models and reporting special cases.
Both authors are funded by the EPSRC grant ``\emph{Informing Future Numerical Standards by Determining Features of Non-Standard Mathematical Hardware}'', ref. UKRI151.

\bibliographystyle{ACM-Reference-Format}
\bibliography{references}

@String{pub-IEEE                = "Institute of Electrical and Electronics Engineers"}

@String{pub-IEEE:adr            = "Piscataway, NJ, USA"}

@String{j-IEEE-TC              = "IEEE Trans. Comput."}

@String{j-SISC                  = "SIAM J. Sci. Comput."}

@String{j-AN                    = "Acta Numerica"}

@String{j-ACM-TOMS              = "ACM Trans. Math. Softw."}

@Misc{amd25,
  author = "AMD",
  title = "Datasheet: {AMD} Instrinct {MI355X GPU}",
  year = 2025,
  url = "https://www.amd.com/content/dam/amd/en/documents/instinct-tech-docs/product-briefs/amd-instinct-mi355x-gpu-brochure.pdf",
  urldate = "2025-07-10"
}

@Article{bhlm20,
author = "Blanchard, Pierre and Higham, Nicholas J. and Lopez, Florent and Mary, Theo and Pranesh, Srikara",
title = "Mixed Precision Block Fused Multiply-Add: Error Analysis and Application to {GPU} Tensor Cores",
journal = "SIAM Journal on Scientific Computing",
volume = 42,
number = 3,
pages = "C124-C141",
year = 2020,
doi = "10.1137/19M1289546"
}

@InProceedings{dgba25,
  author = "Dongarra, Jack and Gunnels, John and Bayraktar, Harun and Haidar, Azzam and Ernst, Dan",
  booktitle = "2025 IEEE High Performance Extreme Computing Conference (HPEC)",
  title = "Accelerating Supercomputing: {AI}-Hardware-Driven Innovation for Speed and Efficiency",
  year = 2025,
  pages = "1-7",
  doi = "10.1109/HPEC67600.2025.11196413"
  }

@Article{fami23,
  author = "Fasi, Massimiliano and Mikaitis, Mantas",
  title = "{CPFloat}: A {C} Library for Simulating Low-Precision Arithmetic",
  year = 2023,
  publisher = "Association for Computing Machinery",
  address = "New York, NY, USA",
  issn = "0098-3500",
  doi = "10.1145/3585515",
  volume = 49,
  number = 2,
  journal = j-ACM-TOMS,
  month = jun,
  articleno = 18,
  numpages = 32,
  pages = "18:1-18:32"
  }

@Article{fhmp21,
  title = "Numerical behavior of {NVIDIA} tensor cores",
  author = "Fasi, Massimiliano and Higham, Nicholas J. and Mikaitis, Mantas and Pranesh, Srikara",
  journal = "PeerJ Computer Science",
  volume = 7,
  pages = "e330",
  year = 2021,
  publisher = "PeerJ Inc.",
  doi = "10.7717/peerj-cs.330"
}

@Article{hbtd20,
author = "Haidar, Azzam  and Bayraktar, Harun  and Tomov, Stanimire  and Dongarra, Jack  and Higham, Nicholas J.",
title = "Mixed-precision iterative refinement using tensor cores on {GPUs} to accelerate solution of linear systems",
journal = "Proceedings of the Royal Society A: Mathematical, Physical and Engineering Sciences",
volume = 476,
number = 2243,
pages = "20200110",
year = 2020,
doi = "10.1098/rspa.2020.0110"
}

@InProceedings{hcry20,
  author = "Hickmann, Brian and Chen, Jieasheng and Rotzin, Michael and Yang, Andrew and Urbanski, Maciej and Avancha, Sasikanth",
  booktitle = "2020 IEEE 27th Symposium on Computer Arithmetic (ARITH)",
  title = "{Intel Nervana Neural Network Processor-T (NNP-T)} Fused Floating Point Many-Term Dot Product",
  year = 2020,
  pages = "133-136",
  doi = "10.1109/ARITH48897.2020.00029"
  }

@InProceedings{hibr19,
  author = "Hickmann, Brian and Bradford, Dennis",
  booktitle = "2019 IEEE 26th Symposium on Computer Arithmetic (ARITH)",
  title = "Experimental Analysis of Matrix Multiplication Functional Units",
  year = 2019,
  month = oct,
  pages = "116-119",
  doi = "10.1109/ARITH.2019.00031"
  }

@InProceedings{hila04,
  author = "Karl E. Hillesland and Anselmo Lastra",
  title = "{GPU} Floating-Point {Paranoia}",
  booktitle = "ACM Workshop on General-Purpose Computing on Graphics Processors (GP$^2$)",
  year = 2004,
  month = aug,
  address = "Los Angeles, CA, USA",
  publisher = "ACM"
  }

@Article{hima22,
  author = "Nicholas J. Higham and Theo Mary",
  title = "Mixed Precision Algorithms in Numerical Linear Algebra",
  journal = j-AN,
  volume = 31,
  pages = "347-414",
  month = may,
  year = 2022,
  doi = "10.1017/s0962492922000022",
  }

@Book{ieee19,
  title = "{IEEE} Standard for Floating-Point Arithmetic,
           {IEEE} {Std} 754-2019 (revision of {IEEE} Std 754-2008)",
  publisher = pub-IEEE,
  year = 2019,
  month = Jul,
  address = pub-IEEE:adr,
  key = "IEEE Computer Society",
  pages = 82,
  doi = "10.1109/IEEESTD.2019.8766229",
  isbn = "978-0-7381-5752-8"
  }

@TechReport{ieee25,
  title = "Interim Report on Binary Floating-point Formats for Machine Learning",
  year = 2026,
  month = jan,
  note = "Version 3.2.1",
  authors = "{IEEE SA P3109 Working Group}",
  url = "https://github.com/P3109/Public/blob/main/IEEE%20WG%20P3109%20Interim%20Report%20v3.2.1.pdf"
  }

@Misc{inte18,
  author = "{Intel Corporation}",
  title = "{BFLOAT16}---Hardware Numerics Definition",
  howpublished = "Available at
                  \url{https://software.intel.com/en-us/download/bfloat16-hardware-numerics-definition}
                  (accessed 15 July 2020)",
  month = nov,
  year = 2018,
  note = "White paper. Document number 338302-001US."
  }

@InProceedings{kamk19,
  author = "Kaul, Himanshu and Anders, Mark and Mathew, Sanu and Kim, Seongjong and Krishnamurthy, Ram",
  booktitle = "2019 IEEE 26th Symposium on Computer Arithmetic (ARITH)",
  title = "Optimized Fused Floating-Point Many-Term Dot-Product Hardware for Machine Learning Accelerators",
  year = 2019,
  pages = "84-87",
  doi = "10.1109/ARITH.2019.00021"
  }

@InProceedings{khmi25,
  author = "A. Khattak, Faizan and Mikaitis, Mantas",
  booktitle = "2025 IEEE High Performance Extreme Computing Conference (HPEC)",
  title = "Generalized Methodology for Determining Numerical Features of Hardware Floating-Point Matrix Multipliers: Part {I}",
  year = 2025,
  doi = "10.1109/HPEC67600.2025.11196657",
  address = "Wakefield, MA, USA",
  month = oct,
  }

@InProceedings{llfs24,
  author = "Li, Xinyi and Li, Ang and Fang, Bo and Swirydowicz, Katarzyna and Laguna, Ignacio and Gopalakrishnan, Ganesh",
  booktitle = "2024 {IEEE} 24th International Symposium on Cluster, Cloud and Internet Computing ({CCGrid})",
  title = "{FTTN}: Feature-Targeted Testing for Numerical Properties of {NVIDIA} \& {AMD} Matrix Accelerators",
  year = 2024,
  pages = "39-46",
  doi = "10.1109/CCGrid59990.2024.00014"
  }

@Article{mami25,
  author = "Mary, Theo and Mikaitis, Mantas",
  title = "Error Analysis of Matrix Multiplication with Narrow Range Floating-Point Arithmetic",
  journal = j-SISC,
  volume = 47,
  number = 4,
  pages = "B785-B800",
  year = 2025,
  doi = "10.1137/24M1685109"
  }

@InProceedings{mclp18,
  author = "S. {Markidis} and S. W. D. {Chien} and E. {Laure}
            and I. B. {Peng} and J. S. {Vetter}",
  title = "{NVIDIA} Tensor Core Programmability, Performance \& Precision",
  booktitle = "Proceedings of the 32nd IEEE International Parallel and
               Distributed Processing Symposium Workshops",
  year = 2018,
  pages = "522-531",
  month = aug,
  address = "Vancouver, BC, Canada",
  doi = "10.1109/IPDPSW.2018.00091"
  }

@Article{mika24,
  author = "Mantas Mikaitis",
  journal = j-IEEE-TC,
  title = "Monotonicity of Multi-Term Floating-Point Adders", 
  year = 2024,
  month = feb,
  volume = 73,
  number = 6,
  pages = "1531-1543",
  doi = "10.1109/TC.2024.3371783"
  }

@TechReport{modc23,
  author = "Micikevicius, Paulius and Oberman, Stuart and Dubey, Pradeep and
                  Cornea, Marius and Rodriguez, Andres and Bratt, Ian and
                  Grisenthwaite, Richard and Jouppi, Norm and Chou, Chiachen and
                  Huffman, Amber and Schulte, Michael and Wittig, Ralph and
                  Jani, Dharmesh and Deng, Summer",
  title = "{OCP} 8-bit Floating Point Specitication ({OFP8})",
  institution = "Open Compute Project",
  year = 2023,
  month = jun,
  note = "Revision 1.0",
  url = "https://www.opencompute.org/documents/ocp-8-bit-floating-point-specification-ofp8-revision-1-0-2023-12-01-pdf-1"
  }

@Misc{rgsm23,
  author = "Bita Darvish Rouhani and Nitin Garegrat and Tom Savell and Ankit More and Kyung-Nam Han and Ritchie Zhao and Mathew Hall and Jasmine Klar and Eric Chung and Yuan Yu and Michael Schulte and Ralph Wittig and Ian Bratt and Nigel Stephens and Jelena Milanovic and John Brothers and Pradeep Dubey and Marius Cornea and Alexander Heinecke and Andres Rodriguez and Martin Langhammer and Summer Deng and Maxim Naumov and Paulius Micikevicius and Michael Siu and Colin Verrilli",
  title = "{OCP} Microscaling Formats ({MX}) Specification",
  url = "https://www.opencompute.org/documents/ocp-microscaling-formats-mx-v1-0-spec-final-pdf",
  note = "Version 1.0",
  pages = 16,
  month = sep,
  year = 2023
  }

@Misc{nvid17,
  author = "NVIDIA",
  title = "{NVIDIA} {Tesla} {V100 GPU} architecture",
  year = 2017,
  url = "https://images.nvidia.com/content/volta-architecture/pdf/volta-architecture-whitepaper.pdf",
  urldate = "2025-06-23"
}

@misc{25dissectingnvidiablackwellarchitecture,
      title={Dissecting the {NVIDIA Blackwell} Architecture with Microbenchmarks}, 
      author={Aaron Jarmusch and Nathan Graddon and Sunita Chandrasekaran},
      year={2025},
      eprint={2507.10789},
      archivePrefix={arXiv},
      primaryClass={cs.DC},
      url={https://arxiv.org/abs/2507.10789}, 
}

@Misc{deep25,
      title={DeepSeek-V3 Technical Report},
      author={DeepSeek-AI},
      year = 2025,
      month = feb,
      eprint = "2412.19437",
      url = "https://arxiv.org/abs/2412.19437"
      }

@misc{finiteelemnt_tc,
      title={Accelerating High-Order Finite Element Simulations at Extreme Scale with FP64 Tensor Cores}, 
      author={Jiqun Tu and Ian Karlin and John Camier and Veselin Dobrev and Tzanio Kolev and Stefan Henneking and Omar Ghattas},
      year={2026},
      eprint={2603.09038},
      archivePrefix={arXiv},
      primaryClass={cs.DC},
      url={https://arxiv.org/abs/2603.09038}, 
}

@INPROCEEDINGS{tc_app_mri,
  author={Lu, Tianjian and Marin, Thibault and Zhuo, Yue and Chen, Yi-Fan and Ma, Chao},
  booktitle={2020 IEEE High Performance Extreme Computing Conference (HPEC)}, 
  title={Accelerating MRI Reconstruction on TPUs}, 
  year={2020},
  volume={},
  number={},
  pages={1-9},
  doi={10.1109/HPEC43674.2020.9286192}}

@INPROCEEDINGS{rounderr,
  author={Taufer, Michela and Padron, Omar and Saponaro, Philip and Patel, Sandeep},
  booktitle={2010 IEEE International Symposium on Parallel \& Distributed Processing (IPDPS)}, 
  title={Improving numerical reproducibility and stability in large-scale numerical simulations on GPUs}, 
  year={2010},
  volume={},
  number={},
  pages={1-9},
  doi={10.1109/IPDPS.2010.5470481}}

@misc{random_ML_review,
      title={Sources of Irreproducibility in Machine Learning: A Review}, 
      author={Odd Erik Gundersen and Kevin Coakley and Christine Kirkpatrick and Yolanda Gil},
      year={2023},
      eprint={2204.07610},
      archivePrefix={arXiv},
      primaryClass={cs.LG},
      url={https://arxiv.org/abs/2204.07610}, 
}

@inproceedings{inference_LLM_numerical,
 author = {Yuan, Jiayi and Li, Hao and Ding, Xinheng and Xie, Wenya and Li, Yu-Jhe and Zhao, Wentian and Wan, Kun and Shi, Jing and Hu, Xia and Liu, Zirui},
 booktitle = {Advances in Neural Information Processing Systems},
 editor = {D. Belgrave and C. Zhang and H. Lin and R. Pascanu and P. Koniusz and M. Ghassemi and N. Chen},
 pages = {169819--169851},
 publisher = {Curran Associates, Inc.},
 title = {Understanding and Mitigating Numerical Sources of Nondeterminism in LLM Inference},
 url = {https://proceedings.neurips.cc/paper_files/paper/2025/file/f80094a824ba5912d4a2de169c404a40-Paper-Conference.pdf},
 volume = {38},
 year = {2025}
}

@INPROCEEDINGS{tc_sp_2,
  author={Oostrum, Leon and Veenboer, Bram and Rook, Ronald and Brown, Michael and Kruizinga, Pieter and Romein, John W.},
  booktitle={2025 IEEE International Parallel and Distributed Processing Symposium (IPDPS)}, 
  title={The Tensor-Core Beamformer: A High-Speed Signal-Processing Library for Multidisciplinary Use}, 
  year={2025},
  volume={},
  number={},
  pages={582-592},
  doi={10.1109/IPDPS64566.2025.00058}}

@inproceedings{sc_tc_3,
author = {Feng, Boyuan and Wang, Yuke and Chen, Guoyang and Zhang, Weifeng and Xie, Yuan and Ding, Yufei},
title = {EGEMM-TC: accelerating scientific computing on tensor cores with extended precision},
year = {2021},
isbn = {9781450382946},
publisher = {Association for Computing Machinery},
address = {New York, NY, USA},
url = {https://doi.org/10.1145/3437801.3441599},
doi = {10.1145/3437801.3441599},
booktitle = {Proceedings of the 26th ACM SIGPLAN Symposium on Principles and Practice of Parallel Programming},
pages = {278–291},
numpages = {14},
location = {Virtual Event, Republic of Korea},
series = {PPoPP '21}
}

@misc{fft_tc_2,
      title={tcFFT: Accelerating Half-Precision FFT through Tensor Cores}, 
      author={Binrui Li and Shenggan Cheng and James Lin},
      year={2021},
      eprint={2104.11471},
      archivePrefix={arXiv},
      primaryClass={cs.DC},
      url={https://arxiv.org/abs/2104.11471}, 
}

@INPROCEEDINGS{fft_tc_1,
  author={Durrani, Sultan and Chughtai, Muhammad Saad and Hidayetoglu, Mert and Tahir, Rashid and Dakkak, Abdul and Rauchwerger, Lawrence and Zaffar, Fareed and Hwu, Wen-mei},
  booktitle={2021 30th International Conference on Parallel Architectures and Compilation Techniques (PACT)}, 
  title={Accelerating Fourier and Number Theoretic Transforms using Tensor Cores and Warp Shuffles}, 
  year={2021},
  volume={},
  number={},
  pages={345-355},
  doi={10.1109/PACT52795.2021.00032}}

@Misc{nvidia_instrset,
  author = "NVIDIA Corporation",
  title = "{CUDA} {B}inary {U}tilities, Release 13.1",
  year = 2025,
  url = "https://docs.nvidia.com/cuda/pdf/CUDA_Binary_Utilities.pdf",
  urldate = "2025-12-19"
}

@Misc{nvid25b,
  author = "NVIDIA",
  title = "{NVIDIA} {Blackwell} Architecture Technical Brief",
  year = 2025,
  url = "https://resources.nvidia.com/en-us-blackwell-architecture",
  urldate = "2025-07-10"
}

@Article{ooyo22,
  author = "Hiroyuki Ootomo and Rio Yokota",
  title = "Recovering single precision accuracy from Tensor Cores while surpassing the {FP32} theoretical peak performance",
  journal = "The International Journal of High Performance Computing Applications",
  volume = 36,
  number = 4,
  pages = "475-491",
  year = 2022,
  month = jun,
  doi = "10.1177/10943420221090256"
  }

@InProceedings{pili21,
  author = "Pisha, Louis and Ligowski, {\L}ukasz",
  title = "Accelerating Non-Power-Of-{$2$} Size {Fourier} Transforms with
           {GPU} Tensor Cores",
  booktitle = "Proceedings of the 2021 IEEE International Parallel and
                  Distributed Processing Symposium",
  mynote = "(IPDPS)",
  year = 2021,
  pages = "507-516",
  month = may,
  address = "Portland, OR, USA",
  doi = "10.1109/IPDPS49936.2021.00059"
  }

@Techreport{schr81,
  author = "N.~L.~Schryer",
  title = "A Test of a Computer's Floating-Point Arithmetic Unit",
  institution = "AT\&T Bell Laboratories, Murray Hill, NJ",
  number = "Computer Science Technical Report 89",
  type = "Technical Report",
  address = "Murray Hill, NJ 07974",
  year = 1981,
  month = feb,
  day = 4
  }

@InProceedings{tbc12,
  author = "Tan, Xuan You and Boland, David and Constantinides, George",
  editor = "Choy, Oliver C. S. and Cheung, Ray C. C. and Athanas, Peter and Sano, Kentaro",
  title = "{FPGA Paranoia}: Testing Numerical Properties of {FPGA} Floating Point IP-Cores",
  booktitle = "Reconfigurable Computing: Architectures, Tools and Applications",
  year = 2012,
  publisher = "Springer Berlin Heidelberg",
  address = "Berlin, Heidelberg",
  pages = "290--301",
  isbn = "978-3-642-28365-9"
}

@InProceedings{tenc09,
  author = "Tenca, Alexandre F.",
  booktitle = "2009 19th IEEE Symposium on Computer Arithmetic",
  title = "Multi-operand Floating-Point Addition",
  year = 2009,
  pages = "161-168",
  doi = "10.1109/ARITH.2009.27"
  }

@InProceedings{vlpg25,
  author = "Valpey, Benjamin and Li, Xinyi and Pai, Sreepathi and Gopalakrishnan, Ganesh",
  title = "An {SMT} Formalization of Mixed-Precision Matrix Multiplication",
  booktitle = "NASA Formal Methods",
  year = 2025,
  publisher = "Springer Nature Switzerland",
  address = "Cham",
  pages = "360--379"
}

\end{document}